\documentclass[showpacs,amsmath,amssymb,aps,pra,twocolumn]{revtex4}
\usepackage{psfrag}
\usepackage{bm}
\usepackage{amsfonts}
\usepackage{amssymb}
\usepackage{graphicx}
\begin{document}

\title{Quantum electrodynamics of qubits}
\author{Iwo Bialynicki-Birula}\email{birula@cft.edu.pl}
\author{Tomasz Sowi\'nski}\email{tomsow@cft.edu.pl}
\affiliation{Center for Theoretical Physics, Polish Academy of Sciences\\
Al. Lotnik\'ow 32/46, 02-668 Warsaw, Poland}

\begin{abstract}
Systematic description of a spin one-half system endowed with magnetic moment or any other two-level system (qubit) interacting with the quantized electromagnetic field is developed. This description exploits a close analogy between a two-level system and the Dirac electron that comes to light when the two-level system is described within the formalism of second quantization in terms of fermionic creation and annihilation operators. The analogy enables one to introduce all the powerful tools of relativistic QED (albeit in a greatly simplified form). The Feynman diagrams and the propagators turn out to be very useful. In particular, the QED concept of the vacuum polarization finds its close counterpart in the photon scattering off a two level-system leading via the linear response theory to the general formulas for the atomic polarizability and the dynamic single spin susceptibility. To illustrate the usefulness of these methods, we calculate the polarizability and susceptibility up to the fourth order of perturbation theory. These {\em ab initio} calculations resolve some ambiguities concerning the sign prescription and the optical damping that arise in the phenomenological treatment. We also show that the methods used to study two-level systems (qubits) can be extended to many-level systems (qudits). As an example, we describe the interaction with the quantized electromagnetic field of an atom with four relevant states: one S state and three degenerate P states.
\end{abstract}
\maketitle
\section{Introduction}

Two-level quantum systems, called {\em qubits} by Schumacher \cite{bs}, play a fundamental role in quantum information theory. In this context they are usually treated as mathematical objects living in a two-dimensional Hilbert space. In reality, qubits always exist as material objects and we should not forget that they are endowed with concrete physical properties. In this paper we shall deal with two-level systems that interact directly with the electromagnetic field, such as spin one-half particles endowed with magnetic moment or two-level atoms. Thus, our results do not apply to qubits encoded in the polarization states of photons. We shall restrict ourselves in this paper to isolated qubits interacting only with the quantized electromagnetic field. Therefore, the calculated decay rates will include only the spontaneous emission.

A two-level system is the simplest model of a quantum system and yet in the presence of a coupling to the quantized electromagnetic field an exact solution has not been obtained. Even in the simplest case, when the electromagnetic field is restricted to just one mode, the model has been exactly solved only in the rotating-wave approximation by Jaynes and Cummings \cite{jc}. Among the approximate solutions, perturbation theory is still the most universal and effective tool, especially in the world of electromagnetic phenomena.
\begin{figure}
\centering
\includegraphics[scale=0.8]{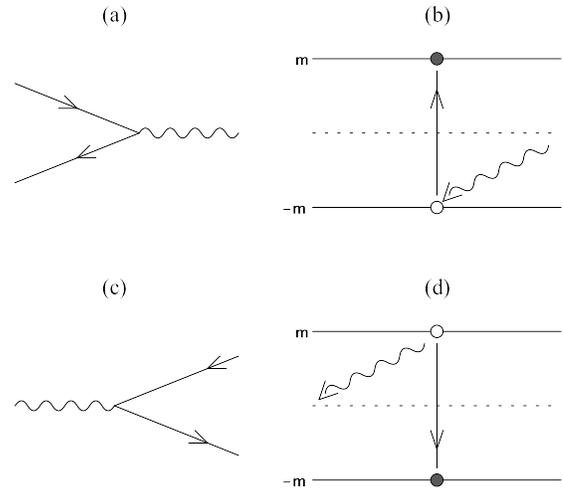}
\caption{Two Feynman diagrams representing the elementary processes and their interpretation in terms of the Dirac-sea picture. The pair creation (a) corresponds to the photon absorption causing a transition (b) of the two-level system from the ground state to the excited state. The electron is moved from the negative energy state (creating a hole) to the positive energy state. The pair annihilation (c) corresponds to the inverse process (d). The electron jumps back from the positive to the negative energy state emitting a photon.}\label{Fig1}
\end{figure}

In the present paper we develop a systematic and complete theory based on an observation that a two-level system can be treated as a relativistic trapped electron. The translational degrees of freedom of such an electron are practically frozen. The only ``degree of freedom'' that remains is the electron's ability to undergo transitions between two discrete energy states. In order to fully unfold the connection between the QED and the theory of two-level systems, we shall perform the second quantization of the standard theory of qubits. The description of two-level systems in terms of creation and annihilation operators has been introduced before (cf., for example, \cite{l}) but no one has exploited the full potential of this formulation. The crucial new element in our formulation is the systematic use of Feynman diagrams. To expose a close analogy with the relativistic theory, including the form of the propagators, we shall choose the energy scale in such a way that the energy levels of the two-level system have opposite signs. In this way, we arrive at a picture of a two-level system that coincides with the Dirac-sea view of quantum electrodynamics. The ground state of the two-level system corresponds to the occupation of the negative energy state, while the excited state corresponds to the occupation of the positive energy state accompanied by a hole in the negative energy sea. The transition between these two states due to the interaction with a photon can be represented by the two elementary Feynman diagrams shown in Fig.~\ref{Fig1}.

There are significant advantages in using the Feynman diagrams and the Feynman propagators associated with these diagrams as compared to the standard perturbation theory used in nonrelativistic quantum mechanics.

First, we never need the formula for the ground state expressed in terms of the noninteracting particles. This is due to the stability of the ground state under the adiabatic switching-on of the interactions. In the Feynman approach the difference between the physical ground state of interacting particles and the ground state of noninteracting particles amounts only to the phase factor corresponding to all disconnected vacuum diagrams \cite{gml,fw}.

\begin{figure}
\centering
\includegraphics{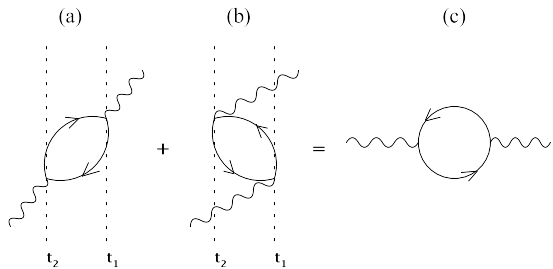}
\caption{Two time orderings in the standard perturbation theory that are combined into one Feynman amplitude represented by one Feynman diagram.}\label{Fig2}
\end{figure}

Second, a single Feynman amplitude combines several terms of the standard perturbation theory since in the Feynman approach all processes that differ only in the time ordering of the vertices are described by one Feynman amplitude (Fig.~\ref{Fig2}). The number of diagrams of the standard perturbation theory that are combined into one Feynman diagram grows exponentially with the number of vertices.

Third, there are many sophisticated tools available to evaluate and analyze Feynman propagators that greatly simplify the calculations and also give a deeper insight into the physical processes described by these propagators. In particular, we shall use the quantum linear response theory to calculate the atomic polarizability and the spin susceptibility from the Feynman propagators. Our formalism is not restricted to two-state systems. It can easily be generalized to many-state systems (qudits) and we analyze as an example a four-state system --- the atomic dipole --- to show that the whole framework can easily be extended to cover this case. The main message of our investigation is that the Feynman description of quantum phenomena, known for its elegance, versatility, and effectiveness in relativistic quantum field theory, also leads to significant simplifications in the theory of qubits. Of course, we are not trying to imply that qubits are relativistic objects. We shall only exploit formal similarities and use many available tools of a relativistic theory. Feynman propagators and Feynman diagrams in our approach should be treated as purely mathematical constructs introduced as a means to streamline and organize perturbation theory. They greatly simplify the calculations but they do not represent any physical objects.

There is a huge number of papers and even a monograph \cite{ae} dealing with the theory of two-level systems and its applications. We believe that the point of view described in this paper will further our understanding of these systems. Our research has been prompted by a recent calculation of the atomic polarizability by Loudon and Barnett \cite{lb}. Our results differ from their results in the fourth order of perturbation theory because they have not taken into account all the necessary corrections. The crossing symmetry of the polarizability, that played an important role in the derivation of the final result by Loudon and Barnett, is automatically satisfied in our formulation. In quantum field theory the crossing relations follow from the analytic properties of the propagators as functions of the energy parameter and from the direct connection between the polarizability and the retarded photon propagator. This connection enabled us to easily calculate the polarizability of a two-level atom and the spin susceptibility in the fourth order of perturbation theory by evaluating the contributions from only a few Feynman diagrams.

Our results clarify certain issues, like the opposite sign versus equal sign prescription or the damping in the ground state, that are still being debated \cite{sted0,bf,sted1,bf1,sted2,mb,bbm}. We show that {\em both sign prescriptions are correct} but they apply to different physical situations. The equal sign prescription is appropriate for the scattering situation when we control the initial and the final photon states. The opposite sign prescription is appropriate in the linear response theory when we control the initial state and also the form of the perturbation but we perform a summation over all final states. Thus, only the opposite-sign convention is appropriate for the calculation of the atomic polarizability. We also show that even though, as stated in \cite{ae}, ``A two-level atom is conceptually the same kind of object as a spin-one-half particle in a magnetic field'', the dynamical properties of these systems are quite different. The differences become significantly different in higher orders of perturbation theory.

Of course, one should keep in mind that our calculations of atomic polarizabilities should not be taken too seriously because the two-level model gives only a very crude description of a real atom. However, for a single spin system, our results are close to reality. The only approximation being made in this case is that the position of the spin is frozen --- the translational degrees of freedom are suppressed.

It has been fully recognized that quantum field theory would, in principle, give unambiguous answers to all such questions but the prevailing opinion that ``there are considerable difficulties associated with the treatment of optical damping in a non-phenomenological manner'' \cite{sted0} discouraged efforts to apply  field-theoretic methods. In this paper we show how to overcome these ``considerable difficulties''. We formulate a theory that is simple because it follows all the rules of a well established theory and it also has an unambiguous interpretation because it is systematically derived from first principles.

In what follows we shall use most of the time a convenient system in units in which $\hslash=1$, $c=1$, and $\mu_0=1$. Of course, in this system of units also $\epsilon_0=1$. More precisely, we express every physical quantity in powers of the meter and $\hslash,c,\mu_0$ (or $\epsilon_0$) and then we drop $\hslash=1$, $c=1$, $\mu_0=1$ and $\epsilon_0$ in the formulas. For example, the Bohr magneton in these units is $\mu_B = 5.847\,10^{-14}\,$m, Tesla is 1T = 5.017$\,10^{15}\,$m$^{-2}$, and the electronvolt is 1eV = 5.068$\,10^{6}\,$m$^{-1}$.

\section{The model Hamiltonian}

The physical system that we shall have in mind is primarily a spinning electron trapped in a spherically symmetric potential subjected to a constant magnetic field and interacting with the quantized electromagnetic field and possibly an external time-varying electromagnetic field. We find it convenient to call this system the electron to stress the analogy with quantum electrodynamics although it is a highly reduced model of an electron. We shall treat in detail the spin system coupled to the electromagnetic field through its magnetic dipole but we shall also extend our analysis to atoms coupled through their electric dipole moments. There are two cases here that must be distinguished: the {\em literal two-level atom} that requires a two-dimensional Hilbert space and an atom with a {\em true electric dipole} moment that requires a four-dimensional Hilbert space that can accommodate the three-dimensional dipole vector.

The Hamiltonian $H=H_0+H_I$ for the spin system in the second-quantized form is
\begin{subequations}\label{hammag}
\begin{align}
H_0&=\int\!d^3r\,{\bm\psi}^\dagger({\bm r})H_0^e{\bm\psi}({\bm r})\nonumber\\
&+\frac{1}{2}\int\!d^3r:\!\left({\bm E}^2({\bm r})+{\bm B}^2({\bm r})\right)\!:\,,\label{hammag0}\\
H_I&=-\mu\int\!d^3r\,{\bm\psi}^\dagger({\bm r}){\bm\sigma}{\bm\psi}({\bm r})\!\cdot\!{\bm B}({\bm r}),\label{hammag1}
\end{align}
\end{subequations}
where $H_0^e$ is the quantum-mechanical Hamiltonian of the electron in the absence of the magnetic coupling and the colons, as usual, denote the normal ordering. We shall assume that the magnetic moment of the electron is coupled to a constant external magnetic field and to the quantized magnetic field. Next, we assume that only the spin degree of freedom is active. Therefore, we can retain only one term in the expansion of the electron field operator
\begin{align}\label{fop}
{\bm\psi}({\bm r})=\chi(r){\bm\psi},
\end{align}
where $\chi({\bm r})$ is a fixed orbital electron wave function assumed to be spherically symmetric. The two-component fermionic operators are ${\bm\psi}^\dagger=(\psi^\dagger_e,\psi^\dagger_g)$ and ${\bm\psi}=(\psi_e,\psi_g)$. Their components create and annihilate the electron in the upper (excited) or lower (ground) energy state. Within this approximation, the Hamiltonian can be rewritten in the form
\begin{subequations}\label{ham}
\begin{align}
H_0&=\mu B_0{\bm\psi}^\dagger\sigma_z{\bm\psi} + \frac{1}{2}\int\!d^3r:\!\left({\bm E}^2({\bm r})+{\bm B}^2({\bm r})\right)\!:\,,\label{ham0}\\
H_I&=-\mu{\bm\psi}^\dagger{\bm\sigma}{\bm\psi} \!\cdot\!\!\int\!d^3r\,\rho(r){\bm B}({\bm r}).\label{ham1}
\end{align}
\end{subequations}
The parameter $\mu$ is the magnetic moment, $B_0$ is the constant magnetic field (pointing in the $z$-direction), and ${\bm\sigma}=(\sigma_x,\sigma_y,\sigma_z)$ are the three Pauli matrices. In the interaction Hamiltonian the magnetic field operator ${\bm B}$ is averaged with the electron distribution function $\rho(r)=\chi^*(r)\chi(r)$ over the region where the trapped electron is localized.

The Hamiltonian $H=H_0+H_I$ conserves the number of electrons. It acts independently in each subspace with a given number of electrons. Since there are just two creation operators in this model, the electronic Fock space is four dimensional. It comprises a one-dimensional zero-particle subspace, a one-dimensional two-particle subspace, and a two-dimensional one-particle subspace spanned by the state vectors $\psi_e^\dagger|0\rangle$ and $\psi_g^\dagger|0\rangle$. This two-dimensional subspace will be our {\em qubit space}. The standard fermionic anticommutation relations
\begin{eqnarray}\label{stand}
\{\psi_i,\psi_j^\dagger\}=\delta_{ij},\;\;\{\psi_i,\psi_j\}=0,\;\;\{\psi_i^\dagger,\psi_j^\dagger\}=0
\end{eqnarray}
imply that the operators ${\bm\psi}^\dagger{\bm\sigma}{\bm\psi}$ annihilate the zero-particle and two-particle sectors, whereas in the qubit space they act as the Pauli matrices. Therefore, in the qubit subspace the Hamiltonian (\ref{ham}) is equivalent to the following one obtained from (\ref{ham}) by replacing all bilinear combinations ${\bm\psi}^\dagger{\sigma}_i{\bm\psi}$ of the operators ${\bm\psi}^\dagger$ and ${\bm\psi}$ by the corresponding Pauli matrices:
\begin{subequations}\label{pauli}
\begin{align}
H_0&=-\mu B_0\sigma_z + \frac{1}{2}\int\!d^3r:\!\left({\bm E}^2({\bm r})+{\bm B}^2({\bm r})\right)\!:\,,\label{pauli1}\\
H_I&=-\mu{\bm\sigma}\!\cdot\!\!\int\!d^3r\,\rho(r){\bm B}({\bm r}).\label{pauli2}
\end{align}
\end{subequations}
To stress the analogy between QED and quantum electrodynamics of two-level systems, from now on we shall denote the energy $\mu B_0$ by the letter $m$.

\subsection{Spin system as a dimensional reduction of QED}

The formulation that employs the electronic creation and annihilation operators will enable us to define new objects --- the propagators --- that do not appear in the standard description of a spin system. The electron propagators, being auxiliary objects without direct physical interpretation, fully deserve the name ``dead wood'', as Dirac \cite{dir} called them. However, a complete formulation of QED (including renormalization) without the propagators would be extremely complicated, if possible at all. We shall show that they are also very useful in the description of two-level systems.

The Hamiltonian (\ref{ham}) acts independently in each sector with a given number of electrons, but the electron creation and annihilation operators cause transitions between these sectors. This leads here, like in full QED, to a greater flexibility of the mathematical formalism and will allow us to introduce objects that are not available in the standard theory of qubits based on the Hamiltonian (\ref{pauli}). Long time ago, the same idea has been successfully applied to the study of the Ising chain \cite{bbp} and that served as an inspiration for the present research. The representation of the spin operators as bilinear expressions of the creation and annihilation operators is the key ingredient of our approach. It enabled us to introduce the fermionic Feynman propagators and to employ the Wick theorem in its most convenient, field-theoretic form that leads directly to standard Feynman diagrams. In contrast, the use of the spin operators as basic variables, does not lead to the Feynman rules in their simplest form known from QED.

In order to better explain the relation between QED and our treatment of two-level systems, let us observe that the Hamiltonian (\ref{ham}) can be obtained by the dimensional reduction from three to zero spatial dimensions. To carry out this reduction, we drop entirely the coordinate dependence and we disregard the integration in the QED Hamiltonian $H_D$ of the Dirac field
\begin{eqnarray}\label{dirac}
H_D=\int\!d^3r\,\left(c\psi^\dagger({\bm r}){\bm\alpha}\!\cdot\!{\bm p}\,\psi({\bm r}) + mc^2\psi^\dagger({\bm r})\beta\psi({\bm r})\right).
\end{eqnarray}
We keep only the mass term and we replace the Dirac field operator $\left(\psi_1({\bm r}),\psi_2({\bm r}),\psi_3({\bm r}),\psi_4({\bm r})\right)$ by the space-independent operators $(\psi_e,\psi_g)$. The operator $\psi_e$ annihilates the particle in the positive energy state and $\psi_g$ annihilates the particle in the negative energy state. The rest energy $m_0c^2$ of the electron is to be identified with $\mu B_0$. Despite these drastic simplifications, we shall still retain the full analogy with quantum electrodynamics. This will enable us to use the highly developed formalism of QED and also to gain deeper insights that go with it.

\subsection{Magnetic dipole Hamiltonian}

Under the assumption that only the spin degree of freedom is active and the orbital part of the electron wave function $\chi(r)$ is fixed and spherically symmetric, only the {\em magnetic dipole} component of the radiation field is coupled to the electron. Therefore, it is most convenient to employ the multipole expansion, i.e. the decomposition of the electromagnetic field into the eigenstates of the angular momentum. Then, the integration of the magnetic field vector with the spherically symmetric distribution in the interaction Hamiltonian (\ref{ham1}) eliminates all multipoles except the magnetic dipole. We present the details of this calculation in the Appendix \ref{a1}. We shall rewrite the Hamiltonian (\ref{finham}) derived there as follows
\begin{align}\label{finham1}
H&=m{\bm\psi}^\dagger\sigma_z{\bm\psi}+\sum_i\int_0^\infty\!dk\,\omega\,c_i^\dagger(k)c_i(k)\nonumber\\
&+{\bm\psi}^\dagger{\bm\sigma}{\bm\psi} \!\cdot\!\!\int_0^\infty\!dk\,g(k){\bm\phi}(k),
\end{align}
where we introduced the dipole vector field ${\bm\phi}(k)$ built from the Cartesian components of the annihilation and creation operators
\begin{eqnarray}\label{phi}
\phi_i(k)=\frac{c_i(k)+c_i^\dagger(k)}{\sqrt{2 k}}.
\end{eqnarray}
The formfactor $g(k)$ is defined in Eq.~(\ref{formf0}) and according to the formula (\ref{formf1}) it is proportional to the Fourier transform ${\tilde\rho}(k)$ of the distribution function $\rho(r)$
\begin{eqnarray}\label{formf}
g(k)=\frac{\mu\,k^2}{\pi\sqrt{3}}{\tilde\rho}(k).
\end{eqnarray}
The normalization condition imposed on $\rho$ requires that ${\tilde\rho}(0)=1$. Therefore, for small values of $k$ the formfactor behaves as $g(k)\approx \mu k^2/\pi\sqrt{3}$. To illustrate this property, let us consider the qubit realized as the spin degree of freedom of a nonrelativistic electron in the ground state of the Coulomb potential. In this case the distribution function $\rho(r)$ and the corresponding formfactor $g(k)$ are
\begin{subequations}\label{gs}
\begin{align}
\rho(r)&=\frac{1}{\pi a_0^3} e^{-2r/a_0},\\
g(k)&=\frac{\mu k^2}{\pi\sqrt{3}}\frac{1}{(1+k^2a_0^2/4)^2},
\end{align}
\end{subequations}
where $a_0$ is the Bohr radius.

The applicability of the model interaction Hamiltonian (\ref{finham1}) extends beyond the simplest case considered here. Should the distribution function $\rho(r)$ be of a more general character or the internal degrees be more complicated, the elimination of higher multipoles could still be justified as an approximation based on the small value of the ratio: atomic size/wavelength.

\subsection{Two-level atom Hamiltonian}

In the case of a literal two-level atom considered by most authors, only {\em one component} of the electromagnetic field is coupled to the atom. Namely, the component that causes transitions between the ground state and one selected excited state. Therefore, it is sufficient to replace the three-component vector ${\bm\phi}(k)$ by a single component $\phi(k)$. In this way, we obtain the standard Hamiltonian for a two-level atom interacting with the quantized electromagnetic field in the form \cite{l,lb}
\begin{align}\label{hamlb}
H&=m\sigma_z+\sum_i\int_0^\infty\!dk\,\omega\,c_i^\dagger(k)c_i(k)\nonumber\\
&+\sigma_x\int_0^\infty\!dk\,{\hat g}(k)\phi(k),
\end{align}
which after the second quantization becomes
\begin{align}\label{hamlb1}
H&=m{\bm\psi}^\dagger\sigma_z{\bm\psi}+\sum_i\int_0^\infty\!dk\,
\omega\,c_i^\dagger(k)c_i(k)\nonumber\\
&+{\bm\psi}^\dagger\sigma_x{\bm\psi}\int_0^\infty\!dk\,{\hat g}(k){\phi}(k),
\end{align}
where the formfactor ${\hat g}(k)$
\begin{eqnarray}\label{formf2}
{\hat g}(k)=\frac{d\,k^2}{\pi\sqrt{3}}{\tilde\kappa}(k).
\end{eqnarray}
is obtained from the formula (\ref{formf}) by replacing the magnetic dipole $\mu$ and its distribution function $\rho$ by the electric dipole $d$ and its distribution function $\kappa$. This natural prescription will be confirmed in the next subsection when we derive the interaction Hamiltonian for a true atomic dipole vector. We place a hat on the symbols of all quantities that refer specifically to two-level atoms to distinguish them from the corresponding quantities for the spin system.

\subsection{Electric dipole Hamiltonian}

The truncation of the atomic Hilbert space to only two dimensions does not allow for the construction of an atomic dipole vector that could be coupled to the electric dipole field. Such a construction can be carried out if we enlarge the Hilbert space of the relevant atomic states to four dimensions. We shall still have only two energy levels but in addition to the ground state we introduce three states corresponding to the degenerate upper level. This is precisely the situation in real atoms if the transitions take place between the ground S state and the three excited P states. The inclusion of all three P states leads to full rotational invariance. Using this specific example we show how to extend our formalism to $N$-level systems. The Hamiltonian $H=H_0+H_I$ expressed in the formalism of second quantization can now be written in the form (cf. Appendix \ref{a1})
\begin{align}\label{1hamel}
H&={\bm\psi}^\dagger {\breve m}{\bm\psi}+\sum_i\int_0^\infty\!dk\,\omega d_i^\dagger(k)d_i(k)\nonumber\\
&+{\bm\psi}^\dagger{\bm\tau}{\bm\psi} \!\cdot\!\!\int_0^\infty\!dk\,{\breve g}(k){\bm\phi}(k),
\end{align}
where we kept the same symbol ${\bm\phi}(k)$ to denote the electromagnetic field because the change from the magnetic dipole field to the electric dipole field does not change any of the mathematical properties of the field ${\bm\phi}(k)$. We introduced four annihilation and four creation operators corresponding to four atomic states. The operators for the ground state and the operators for the excited states in the Cartesian basis are combined into four-dimensional objects ${\bm\psi}=\{\psi_x,\psi_y,\psi_z,\psi_g\}$ and ${\bm\psi}^\dagger=\{\psi_x^\dagger,\psi_y^\dagger,\psi_z^\dagger,\psi_g^\dagger\}$. They obey the fermionic anticommutation relations (\ref{stand}). The matrices ${\breve m}$ and ${\bm\tau}$ are defined in Eqs.~(\ref{mat}). The derivation in Appendix \ref{a1} of the formula for the formfactor function ${\breve g}(k)$ gives the precise meaning to the dipole moment $d$ of the atomic transition and the dipole distribution function $\kappa(r)$ and its transform ${\breve\kappa}(k)$.
\begin{align}\label{wg}
{\breve g}(k)=\frac{d\,k^2}{\pi\sqrt{3}}{\breve\kappa}(k).
\end{align}
Since for small values of $k$ we have $j_1(kr)\approx kr/3$, the function ${\breve\kappa}(k)$ has the same normalization as ${\tilde\rho}(k)$ --- it approaches 1, when $k\to 0$. In particular, for the P-S transitions in the hydrogen atom we obtain
\begin{subequations}\label{khyd}
\begin{align}
\kappa(r)&=\frac{er^2}{4\pi a_0^4d\sqrt{6}}\exp\left(-\frac{3r}{2a_0}\right),\\
{\breve g}(k)&=\frac{d\,k^2}{\pi\sqrt{3}}\frac{1}{\left(1+4k^2a_0^2/9\right)^3}.\\
d&=\frac{2^{15/2}ea_0}{3^5}.
\end{align}
\end{subequations}

\subsection{Conservation of angular momentum}

The interaction Hamiltonian for the spin system is invariant under all rotations since it is a scalar products of two vectors. However, the full Hamiltonian is invariant only under rotations around the $z$ axis since the free fermion Hamiltonian (\ref{finham1}) contains the $z$ component of the vector ${\bm\sigma}$. The physical origin of the symmetry breaking is the external magnetic field $B_0$ fixed along the $z$-axis. It splits the energy levels of the magnetic dipole and breaks the full rotational invariance. In contrast, the Hamiltonian for the electric dipole is invariant under the full rotation group. This invariance is possible because the Coulomb potential of the hydrogenic atom is rotationally symmetric and we have included all three components of the excited P state. These components form a vector representation of the rotation group.

The invariance of the Hamiltonian implies the commutativity of the angular momentum operator $M_z$ with $H$ leading to the conservation of the $M_z$ in both cases. The angular momentum operators for the spin system and for the electric dipole are
\begin{align}\label{mtot}
M_i&=\frac{1}{2}{\bm\psi}^\dagger\sigma_i{\bm\psi}
-i\int_0^\infty\!\!\!dk\,\epsilon_{ijk}c^\dagger_j(k)c_k(k),\\
{\breve M}_i&={\bm\psi}^\dagger(k){\bm\psi}(k)
-i\int_0^\infty\!\!\!dk\,\epsilon_{ijk}d^\dagger_j(k)d_k(k),
\end{align}
where the spin-one matrices $s_i$ with elements $(s_i)_{jk}=-i\epsilon_{ijk}$ act in the subspace of excited states. Conservation of angular momentum during interaction becomes obvious when the angular momentum operator and interaction Hamiltonian are written in the angular momentum basis. We shall use the spin system to illustrate these properties. Let us construct the components of the magnetic dipole field ${\phi}_\pm(k)$ and ${\phi}_0(k)$ from the annihilation and creation operators of photons with the definite angular momentum $M_z=\pm 1,0$ introduced in the Appendix
\begin{subequations}
\begin{align}\label{phif}
&{\phi}_+(k)=\frac{c_-(k)-c_+^\dagger(k)}{\sqrt{2 k}},\\
&{\phi}_-(k)=\frac{c_-^\dagger(k)-c_+(k)}{\sqrt{2 k}}={\phi}_+^\dagger(k),\\
&{\phi}_0(k)=\frac{c_0(k)+c_0^\dagger(k)}{\sqrt{2 k}}.
\end{align}
\end{subequations}
The operators $M_z$ and $H_I$ take now the form
\begin{equation}\label{finmz}
M_z=\frac{1}{2}\psi^\dagger\sigma_z\psi +\int_0^\infty\!\!\!\!dk\,\left[c^\dagger_{+}(k)c_{+}(k)-c^\dagger_{-}(k)c_{-}(k)\right],
\end{equation}
\begin{align}\label{finham2}
H_I&=\psi^\dagger{\sigma}_+\psi\int_0^\infty\!\!\!dk\,g(k){\phi}_-(k)
+\psi^\dagger{\sigma}_-\psi\int_0^\infty\!\!\!dk\,g(k){\phi}_+(k)\nonumber\\
&+\psi^\dagger{\sigma}_z\psi\int_0^\infty\!dk\,g(k){\phi}_0(k),
\end{align}
where
\begin{align}\label{angsig}
\sigma_+=\frac{\sigma_x+i\sigma_y}{\sqrt{2}},\;\;\;\sigma_-=\frac{\sigma_x-i\sigma_y}{\sqrt{2}}.
\end{align}
The field ${\phi}_+(k)$ coupled to $\sigma_-$ annihilates the photon with $M_z=-1$ or creates the photon with $M_z=1$. Thus, it increases the angular momentum by one unit. The field ${\phi}_-(k)$ coupled to $\sigma_+$ decreases the angular momentum by one unit. Each term in the Hamiltonian (\ref{finham2}) conserves angular momentum. For example, when $\sigma_+$ transfers the electron from the ground state to the excited state increasing its angular momentum by one (the first term), the angular momentum of the electromagnetic field decreases by one unit. Similar analysis can be carried out for the electric dipole. Of course, for the literal two-level atom there is no invariance under rotation because only one angular momentum state of the photon interacts with the atom. Hence, only one component of the electronic P state (and not all three) can be excited.

\subsection{Time-reversal invariance}

Both theories, describing the spin and the two-level atom, are invariant under the time reversal. This invariance can be proven directly but it also follows from the fact that our models are obtained by the dimensional reduction from QED which has this property. Time-reversal invariance is an important requirement to obtain a correct description of the optical damping, as stressed in Ref.~\cite{sted1}. In what follows we shall make use of this invariance. Under the time reversal the signs of the frequency and angular momentum are reversed. Therefore, there is no need to calculate the photon propagator for the negative values of $M_z$ for the spin system because they can be obtained from those for the positive values by reversing the sign of the frequency. When the results are the same for positive and negative values of  $M_z$, as is the case for the atomic system, time-reversal invariance means that the photon propagator is an even function of the frequency. The conservation of angular momentum and time-reversal invariance simplify the calculations since they reduce the number of Feynman integrals that are to be evaluated.

\section{Propagators and the $S$ matrix}

All transition amplitudes can be expressed in terms of Feynman propagators --- the expectation values in the ground state of the time-ordered products of field operators. Since we shall be working within perturbation theory, the most useful representation of the propagators is the one that is based on the perturbative expansion of the $S$ matrix. The relevant formula for the $S$ matrix is the following standard expansion into the time-ordered products of the interaction Hamiltonians \cite{dys}:
\begin{align}\label{sop}
S&=T\exp\left(-i\int\!dt\,H_I(t)\right)\nonumber\\
&\equiv\sum_{n=0}^\infty\frac{(-i)^n}{n!}\int\!dt_1\cdots\int\!dt_nT\left[H_I(t_1)\cdots H_I(t_n)\right].
\end{align}
The interaction Hamiltonian in this formula is taken in the Dirac picture. We shall introduce all the necessary theoretical tools starting with the spin system but later extending them to atoms by making obvious modifications. We will find it expedient, even though it is not necessary since there are no infinities, to perform the mass renormalization. This amounts, exactly like in QED, to adding the mass-correction term $\delta m\psi^\dagger\sigma_z\psi$ to the free Hamiltonian and subtracting the same term from the interaction Hamiltonian. In our case, the freedom of choosing $\delta m$ can be viewed as a mechanism to improve the convergence of perturbation theory. After the mass renormalization, the free Hamiltonian and interaction Hamiltonian in the Dirac picture become
\begin{align}\label{freeham}
&H_0=(m_0+\delta m){\bm\psi}^\dagger\sigma_z{\bm\psi}\nonumber\\
&+ \frac{1}{2}\int_0^\infty\!dk:\!\left({\bm\pi}^2(k)+k^2\phi^2(k)\right)\!:,
\end{align}
\begin{align}\label{intham}
&H_I(t)=e^{iH_0t}H_Ie^{-iH_0t}\nonumber\\
&={\bm\psi}^\dagger(t){\bm\sigma}{\bm\psi}(t)\!\cdot\!\!\int_0^\infty\!dk\,g(k){\bm\phi}(k,t) -\delta m{\bm\psi}^\dagger(t)\sigma_z{\bm\psi}(t),
\end{align}
where ${\bm\pi}(k)$ is the canonically conjugate momentum
\begin{align}\label{canmom}
{\bm\pi}(k)=-i\sqrt{\frac{k}{2}}\left({\bm c}(k)-{\bm c}^\dagger(k)\right).
\end{align}
The time dependence of the operators ${\bm\psi}(t),{\bm\psi}^\dagger(t)$, and ${\bm\phi}(k,t)$ is determined by the renormalized fermionic Hamiltonian (\ref{freeham}) and it has the following form:
\begin{eqnarray}\label{fintpic}
{\bm\psi}(t)&=&\left(\begin{array}{c}\psi_e e^{-imt}\\\psi_ge^{imt}\end{array}\right),\nonumber\\
{\bm\psi}^\dagger(t)&=&\left(\psi_e^\dagger e^{imt},\psi^\dagger_ge^{-imt}\right),\\
\end{eqnarray}
where $m=m_0+\delta m$. The time dependence of the field ${\bm\phi}(k,t)$ is
\begin{eqnarray}\label{bintpic}
{\bm\phi}(k,t)=\frac{{\bm c}(k)e^{-i\omega t}+{\bm c}^\dagger(k)e^{i\omega t}}{\sqrt{2 k}}.
\end{eqnarray}
Note that due to our normalization, the electromagnetic field operators ${\bm\phi}(k,t)$ and ${\bm\pi}(k',t)={\dot{\bm\phi}}(k',t)$ satisfy the equal-time canonical commutation relations
\begin{eqnarray}\label{ccr0}
[\phi_i(k,t),\pi_j(k',t)]=i\delta_{ij}\delta(k-k').
\end{eqnarray}

In order to describe the interacting system, we need the propagators defined in terms of the field operators ${\bm\Psi}(t),{\bm\Psi}^\dagger(t)$, and ${\bm\Phi}(k,t)$ evaluated in the Heisenberg picture. We shall use lower case and upper case letters to keep the distinction between the Dirac (interaction) picture and the Heisenberg picture operators. The Heisenberg picture operators obey the following equations of motion:
\begin{subequations}\label{heqm}
\begin{align}
(i\partial_t-m_0\sigma_z)\Psi(t)&=\int_0^\infty\!dk g(k){\bm\sigma}\!\cdot\!{\bm\Phi}(k,t)\Psi(t),\label{heqm1}\\
(\partial_t^2+k^2){\bm\Phi}(k,t)&=-g(k)\Psi^\dagger(t){\bm\sigma}\Psi(t)\label{heqm2}.
\end{align}
\end{subequations}
The canonical equal-time commutation relations of the Heisenberg operators are the same as their free counterparts
\begin{subequations}\label{ccr}
\begin{align}
\left\{\Psi_\alpha(t),\Psi_\beta^\dagger(t)\right\}&=\delta_{\alpha\beta},\label{ccr1}\\
\left[\Phi_i(k,t),{\dot\Phi}_j(k',t)\right]&=i\delta_{ij}\delta(k-k')\label{ccr2}.
\end{align}
\end{subequations}
All remaining commutators or anticommutators vanish.

The perturbation expansion of the propagators can be obtained from the following formula \cite{gml,bd} by expanding the time-ordered exponential function into a power series according to Eq.~(\ref{sop}):
\begin{widetext}
\begin{align}\label{sprop}
\langle G|T\big[\Psi(t_1)\cdots\Psi(t_i)\Psi^\dagger(t_1')&\cdots\Psi^\dagger(t_i') \Phi(k_1,t_1'')\cdots\Phi(k_l,t_l'')\big]|G\rangle\nonumber\\
&=\frac{\langle g|T\left[\psi(t_1)\cdots\psi(t_i)\psi^\dagger(t_1')\cdots\psi^\dagger(t_i') \phi(k_1,t_1'')\cdots \phi(k_l,t_l'')\exp\left(-i\int\!dt\,H_I(t)\right)\right]|g\rangle}{\langle g|T\exp\left(-i\int\!dt\,H_I(t)\right)|g\rangle}.
\end{align}
\end{widetext}
We have omitted here all indices leaving only the dependence on time and on the wave vector. The operators on the left hand side of this equation are in the Heisenberg picture while those on the right hand side are all in the Dirac picture. In this formula $|G\rangle$ denotes the true ground state of the interacting system and $|g\rangle$ denotes the ground state of the free Hamiltonian $H_0$. In the state $|g\rangle$ there are no photons and the negative energy state of the electron is occupied. The advantage of using this fundamental result, already mentioned in the Introduction, is that the detailed knowledge of the ground state $|G\rangle$ is not needed. The difference between the state vectors $|G\rangle$ and $|g\rangle$ is just a phase factor and the denominator in the formula (\ref{sprop}) representing the contributions from all disconnected vacuum diagrams takes care of that.

\section{Feynman diagrams and Feynman rules}

In order to derive the Feynman rules that connect the Feynman diagrams with the corresponding transition amplitudes we start, as in QED, from the free field operators. The time evolution of these operators is given by Eqs.~(\ref{fintpic}) and (\ref{bintpic}).

The basic ingredients of the Feynman formulation of QED are the free one-electron propagator $S_F$ and one-photon propagator $D_F$. In our model they are defined as follows:
\begin{align}
S_{F\alpha\beta}(t-t')&=-i\langle g|T\left(\psi_\alpha(t)\psi^\dagger_\beta(t')\right)|g\rangle,\label{elprop}\\
\!\!D_{Fij}(k,k',t-t')&=-i\langle g|T\left(\phi_i(k,t)\phi_j(k',t')\right)|g\rangle,\label{phprop}
\end{align}
where $|g\rangle$ is the ground state of the system without interaction. We have introduced the photon propagator only for those photons that are coupled to the electron.

\subsection{Free electron propagators}

The free electron propagator is easily evaluated with the use of Eqs.~(\ref{fintpic}) taking into account that the only nonvanishing matrix elements of the bilinear product of the creation and annihilation operators are $\langle g|\psi_e\psi_e^\dagger|g\rangle=1$ and $\langle g|\psi_g^\dagger\psi_g|g\rangle=1$. Therefore, we obtain
\begin{align}\label{eprop0}
&iS_{F\alpha\beta}(t-t')\nonumber\\
=&\theta(t-t')\langle
g|\psi_\alpha(t)\psi^\dagger_\beta(t')|g\rangle\!
-\theta(t'\!-t)\langle g|\psi^\dagger_\beta(t')\psi_\alpha(t)|g\rangle\nonumber\\
=&\theta(t-t'){\mathbb
P}_{e\alpha\beta}e^{-im_e(t-t')}-\theta(t'-t){\mathbb
P}_{g\alpha\beta}e^{-im_g(t-t')},
\end{align}
where ${\mathbb P}_e=(1+\sigma_z)/2$ and ${\mathbb
P}_g=(1-\sigma_z)/2$ are the projection matrices on the upper and lower energy states, respectively. For the spin system and the two-level atom we have $m_e=m$ and $m_g=-m$. However, for the dipole atom these two parameters will be independent. The final result can be expressed in matrix notation (omitting the indices $\alpha$ and $\beta$)
as the following Fourier integral:
\begin{eqnarray}\label{eprop}
{S}_{F}(t-t')=\int_{-\infty}^\infty\!\frac{d p_0}{2\pi}{S}_{F}(p_0)e^{-ip_0(t-t')},
\end{eqnarray}
where ${S}_{F}(p_0)$ has the form
\begin{subequations}\label{eprop1}
\begin{align}
{S}_{F}(p_0)&=\frac{{\mathbb P}_e}{p_0-m_e+i\epsilon}+\frac{{\mathbb
P}_g}{p_0-m_g-i\epsilon}\label{epropa}\\
&=\frac{\sigma_z}{p_0{\sigma}_z-m+i\epsilon}\label{epropb}\\
&=\frac{1}{p_0-(m-i\epsilon){\sigma}_z}\label{epropc}.
\end{align}
\end{subequations}
The formula (\ref{epropa}) holds also for the atomic dipole when the excited states form a subspace. In what follows we shall use the same symbols ${\mathbb P}_e$ and ${\mathbb P}_g$ to denote the projectors in all three cases. It will be clear from the context, whether ${\mathbb P}_e$ projects on the one-dimensional subspace (spin and two-level atom) or on the three dimensional subspace (atomic dipole). As compared with the Fourier transform of the electron propagator in the relativistic theory $1/(\gamma\cdot p -m+i\epsilon)$, the two-level propagator (\ref{eprop1}) lacks the spatial part of the momentum vector and has the Pauli $\sigma_z$ matrix instead of $\gamma_0$. The presence of $\sigma_z$ in the numerator in Eq.~(\ref{epropa}) reflects the fact that we work with ${\bm\psi}^\dagger$ instead of $\bar{\bm\psi}={\bm\psi}\gamma_0$. We shall use the same symbols to denote the propagators and their Fourier transforms. The arguments will always indicate which is the case.

\subsection{Free photon propagators}

The free photon propagator is
\begin{align}\label{pprop0}
D_{Fij}(k,k',t-t')&=-i\theta(t-t')\langle g|\phi_i(k,t)\phi_j(k',t')|g\rangle\nonumber\\
&-i\theta(t'-t)\langle g|\phi_i(k',t')\phi_j(k,t)|g\rangle\nonumber\\
&=-i\frac{\delta_{ij}\delta(k-k')}{2k}e^{-i\omega|t-t'|}.
\end{align}
We shall also need its Fourier representation
\begin{align}
D_{Fij}(k,k',t-t')=\int\!\frac{dk_0}{2\pi}D_{Fij}(k,k',k_0)e^{-ik_0(t-t')},
\end{align}
where
\begin{subequations}\label{pprop}
\begin{align}
&D_{Fij}(k,k',k_0)=\frac{\delta_{ij}\delta(k-k')}{k_0^2-k^2+i\epsilon}\label{ppropa}\\
&=\frac{\delta_{ij}\delta(k-k')}{2k}\left(\frac{1}{k_0-k+i\epsilon}-\frac{1}{k_0+k-i\epsilon}\right)\label{ppropb}.
\end{align}
\end{subequations}
All Feynman amplitudes can be constructed from the electron propagator (\ref{eprop}), the photon propagator (\ref{pprop}), the vertex, and the mass insertion following the same general rules as in QED. The starting point is the definition (\ref{sprop}) of a general propagator. In the $n$-th order of perturbation theory the contribution to the propagator is expressed as an expectation value of the time-ordered product of operators ${\bm\psi}, {\bm\psi}^\dagger$, and ${\bm\phi}$ integrated over $n$ time variables. In our model, as in the standard QED, all these expectation values can be evaluated with the help of the Wick theorem (cf., for example, \cite{bd,iz}). The only difference in applying this theorem is, in contrast to QED, that we have not interchanged the creation and annihilation operators for the negative energy state. Calling the electron in the ground state an antiparticle would stretch the analogy with QED too far. Therefore, in our case the normal ordering means that all operators $\psi^\dagger_e$ and $\psi_g$ stand to the left of all operators $\psi^\dagger_g$ and $\psi_e$.

\subsection{Feynman rules}

\begin{figure}
\centering \vspace{0.5cm}
\includegraphics[scale=1.2]{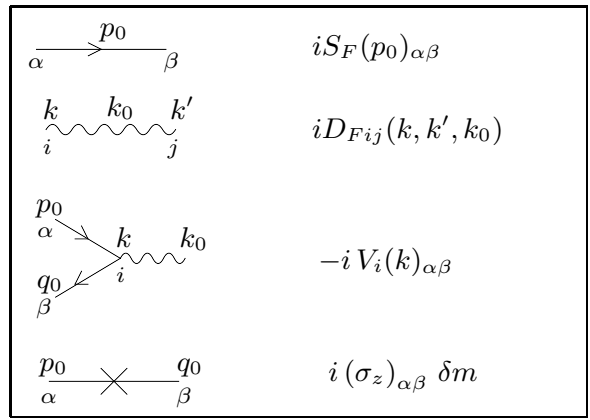}
\caption{Feynman rules. For clarity, we have written explicitly all indices. }\label{Fig3}
\end{figure}

The scattering amplitudes in QED are commonly evaluated in momentum representation. In our case, the transformation to momentum representation means the transformation from the time domain to the frequency domain. The Feynman rules in the frequency domain are obtained by substituting everywhere the free electron propagators and photon propagators in the form of the Fourier integrals (\ref{eprop}) and (\ref{pprop}). Next, in the $n$-th order of perturbation theory we perform $n$ time integrations. Finally, we take the inverse Fourier transforms with respect to all remaining time arguments of the propagator (\ref{sprop}). These operations lead to the following Feynman rules:

\begin{itemize}
\item Each electron line corresponds to the Fourier transform of the electron propagator and is represented by $iS_F(p_0)$.
\item Each photon line corresponds to the Fourier transform of the photon propagator and is represented by $iD_{Fij}(k,k',k_0)$.
\item Each vertex is depicted by two electron lines and the photon line meeting at one point. It is represented by $-iV_i(k)=-ig(k)\sigma_i$. The energy conservation at each vertex results in the appearance of $2\pi\delta(p_0-q_0-k_0)$.
\item Each mass insertion is depicted by a cross where two electron lines meet. It is represented by $i\delta m\sigma_z$.  The energy conservation at each mass insertion results in the appearance of $2\pi\delta(p_0-q_0)$.
\item All $2\times 2$ matrices corresponding to electron propagators are multiplied in the order indicated by the arrows on the diagram.
\item Each closed electronic loop brings in a minus sign and a trace over the matrix indices.
\item There is a summation over all repeated vector indices and an integration over all repeated values of the length of the wave vector.
\item There is one integration over the energy variable for each closed loop, accompanied by the division by $2\pi$.
\end{itemize}
These rules are summarized in Fig.~\ref{Fig3}. Calculations of the lowest order radiative corrections to the electron and photon propagators based on these rules are presented in Sections \ref{sep} and \ref{spp}.

In the case of the two-level atom the only changes in the Feynman rules as compared to the case of the spin system is that the elementary vertex is represented just by $-ig(k)\sigma_x$ and the photon propagator has no indices. In the case of the atom with an electric dipole the free photon propagator retains its form (\ref{pprop}). The free electron propagator must be taken in the general form (\ref{epropa})
\begin{align}\label{elpropgen}
S_F(p_0)=\frac{{\mathbb P}_e}{p_0-m_e+i\epsilon}+\frac{{\mathbb
P}_g}{p_0-m_g-i\epsilon}
\end{align}
and at each vertex the matrices ${\bm\sigma}$ must be replaced by the matrices ${\bm\tau}$.

\section{Radiative corrections}

Owing to the absence of the space components of momentum vectors, the calculation of radiative corrections is much simpler here than in the full-fledged QED. There is no need to combine denominators \'a la Feynman and Schwinger. All integrations with respect to the loop variables $p_0, k_0$ etc. can be evaluated analytically by the residue method {\em in any order of perturbation theory}. At the end we will be left only with the integrals over the wave vectors of photons weighted with $g^2(k)$. Of course, those integrals cannot be evaluated if the function $g(k)$ is not specified.

\begin{figure}
\centering \vspace{0.5cm}
\includegraphics[scale=0.8]{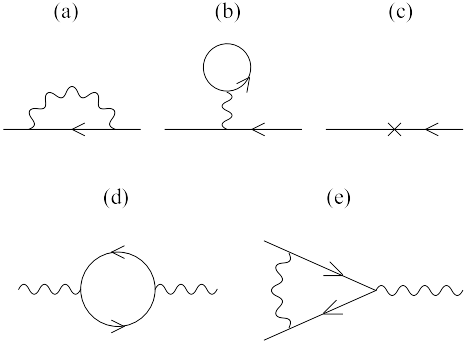}
\caption{Feynman diagrams representing the lowest-order radiative corrections to the electron propagator, photon propagator, and the vertex part.}\label{Fig4}
\end{figure}

\begin{figure}
\centering \vspace{0.5cm}
\includegraphics[scale=0.8]{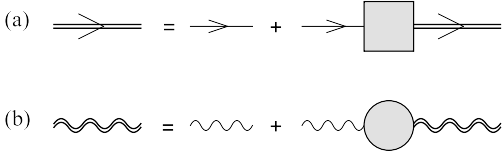}
\caption{Graphical representation of the relationship between the propagators and the corresponding self-energy parts. The double-lines represent full propagators and the gray box and circle represent the self-energy parts.}\label{Fig5}
\end{figure}

In order to explain how the calculations are done, let us consider an integral represented by an arbitrary Feynman diagram. The integrand is a product of electron and photon propagators. To perform all the integrations with respect to the loop variables, one may choose the electron propagator in the form (\ref{epropa}) and use the photon propagator in the form (\ref{ppropb}). The numerator of the integrand corresponding to each Feynman diagram is a polynomial in the integration variables. The denominator is a product of first-order polynomials in the integration variables, each factor leading to a simple pole. All integrations can easily be done by the standard residue method. Note that after each successive integration the integrand retains its rational form. Therefore, it will continue to be amenable to the same treatment as during the first integration. Alternatively, we may choose the interaction Hamiltonian in the angular momentum basis (\ref{finham2}). The following algebraic properties of the matrices $\sigma_\pm$ are then very useful:
\begin{subequations}\label{algprop}
\begin{eqnarray}
\sigma_+^2=0=\sigma_-^2,\;\;\sigma_+\sigma_-=2{\mathbb P}_e,\;\;\sigma_-\sigma_+=2{\mathbb P}_g,\\
\sum_n\sigma_n M \sigma_n = \sigma_+ M\sigma_- + \sigma_- M\sigma_+ +\sigma_z M\sigma_z,\\
\sum_n\!\sigma_n {\mathbb P}_e \sigma_n = {\mathbb P}_e+2{\mathbb P}_g,\;
\sum_n\!\sigma_n {\mathbb P}_g \sigma_n = {\mathbb P}_g+2{\mathbb P}_e,
\end{eqnarray}
\end{subequations}
where $M$ is an arbitrary matrix. With their help, and using the anticommutativity of $\sigma_\pm$ with $\sigma_z$, we can reduce every Feynman integral to to a very simple form.

In the case of a two-level atom the calculations are simpler than in the case of the spin system. Due to the appearance of only the $\sigma_x$ matrix in all vertices, the matrix algebra is almost trivial. In each integrand we can bring up front all $\sigma_x$ matrices using the relations $\sigma_x\sigma_z=-\sigma_z\sigma_x$ and $\sigma_x^2=1$. Therefore, each time we interchange the order of $\sigma_x$ and $\sigma_z$ in the electron propagator the sign of $\sigma_z$ must be reversed. Since there will be an even number of vertices in all the diagrams under consideration, the matrices $\sigma_x$ will disappear completely and we will be left with a diagonal matrix that contains only the matrices $\sigma_z$. The trace of such an expression is the sum of the terms corresponding to the eigenvalues $\pm 1$ of $\sigma_z$.

In the case of the electric dipole, the following algebraic properties of the ${\bm\tau}$ matrices:
\begin{eqnarray}\label{algprop1}
\tau_i{\mathbb P}_g={\mathbb P}_e\tau_i,\;\;\;\tau_i{\mathbb P}_e={\mathbb P}_g\tau_i,\;\;
\sum_n\tau_n\tau_n={\mathbb P}_e+3{\mathbb P}_g,
\end{eqnarray}
used in conjunction with the general form (\ref{epropa}) of the free electron propagator greatly reduce the number of integrals that are to be evaluated.

We shall show how these rules work in practice by calculating radiative corrections to the electron and photon propagators. The procedure employed very often in QED relates the full electron and photon propagators to the self-energy parts. This procedure enables one to go beyond the simplest version of perturbation theory and sum up an infinite (geometric) series. The self-energy is the sum of contributions from strongly connected diagrams, i.e. the diagrams that cannot be disconnected by cutting only one line. The relations between the full propagators and the self-energy parts are shown schematically in Fig.~\ref{Fig5}.

\section{Electron propagator}\label{sep}

In the case of the electron propagator $G_F(p_0)$ the relation between the propagator and the self-energy part $\Sigma(p_0)$, illustrated in Fig.~\ref{Fig5}a, reads
\begin{eqnarray}\label{g}
G_F(p_0)=S_F(p_0)+S_F(p_0)\Sigma(p_0)G_F(p_0).
\end{eqnarray}
All three objects that appear in this equation are $2\times 2$ matrices. The iterative solution of Eq.~(\ref{g}) that shows explicitly the relation between the propagator and the self-energy part, is
\begin{align}\label{git}
G_F(p_0)&=S_F(p_0)+S_F(p_0)\Sigma(p_0)S_F(p_0)\nonumber\\
&+S_F(p_0)\Sigma(p_0)S_F(p_0)\Sigma(p_0)S_F(p_0)+\dots.
\end{align}
This formal geometric series can be summed up to the following compact form:
\begin{eqnarray}\label{g1}
G_F(p_0)=\frac{1}{S_F^{-1}(p_0)-\Sigma(p_0)},
\end{eqnarray}
where the inverse is to be understood as the inverse of a matrix. The series (\ref{git}) without resummation is meaningless because it is divergent when $p_0\approx m$.

The radiative corrections to the electron propagator in the second order of perturbation theory are represented by the three Feynman diagrams (a)--(c) shown in Fig.~\ref{Fig4}. The self-energy parts in this order for the spin system ${\Sigma}^{(2)}(p_0)$, for the two-level atom ${\hat\Sigma}^{(2)}(p_0)$, and for the dipole atom ${\breve\Sigma}^{(2)}(p_0)$, constructed according to the rules stated in the previous section, have the form

\begin{widetext}
\begin{subequations}
\begin{align}\label{sigma}
\Sigma^{(2)}(p_0)&=\Sigma^{(2a)}(p_0)+\Sigma^{(2b)}(p_0)+\Sigma^{(2c)}(p_0)= i\int_{-\infty}^{\infty}\!\frac{dk_0}{2\pi}\sum_{i}\int_0^\infty\!\!dk\sum_{j}\int_0^\infty\!\!dk'
V_i(k)S_F(p_0+k_0)V_j(k')D_{Fij}(k,k',k_0)\nonumber\\
&-i\int_{-\infty}^{\infty}\!\frac{dp_0}{2\pi}\sum_{i}\int_0^\infty\!\!dk\sum_{j}\int_0^\infty\!\!dk'
{\mathrm Tr}\{V_i(k)S_F(p_0)\}V_j(k')D_{Fij}(k,k',0)-\delta m\,\sigma_z,\\
{\hat\Sigma}^{(2)}(p_0)&= {\hat\Sigma}^{(2a)}(p_0)+{\hat\Sigma}^{(2c)}(p_0)= i\int_{-\infty}^{\infty}\!\frac{dk_0}{2\pi}\int_0^\infty\!\!dk\int_0^\infty\!\!dk'
V(k)S_F(p_0+k_0)V(k')D_{F}(k,k',k_0)-\delta{\hat m}\,\sigma_z,\\
{\breve\Sigma}^{(2)}(p_0)&= {\breve\Sigma}^{(2a)}(p_0)+{\breve\Sigma}^{(2c)}(p_0)= i\int_{-\infty}^{\infty}\!\frac{dk_0}{2\pi}\sum_{i}\int_0^\infty\!\!dk\sum_{j}\int_0^\infty\!\!dk'
V_i(k)S_F(p_0+k_0)V_j(k')D_{Fij}(k,k',k_0)-\delta{\breve m}.
\end{align}
\end{subequations}
The tadpole diagram (Fig.~\ref{Fig4}b) does not contribute in the case of the two-level atom and the dipole atom because ${\mathrm Tr}\{\sigma_x S_F(p_0)\}=0$ and ${\mathrm Tr}\{\tau_i S_F(p_0)\}=0$. The analytic expressions for the self-energy parts obtained by the application of the Feynman rules are
\begin{subequations}
\begin{align}\label{sigma1}
&\Sigma^{(2)}(p_0)
=i\sum_{n}\int_0^\infty\!\!dk\,g^2(k)\int_{-\infty}^{\infty}\!\frac{dk_0}{2\pi}
\sigma_n\frac{1}{p_0+k_0-(m-i\epsilon)\sigma_z}\sigma_n\frac{1}{k_0^2-k^2+i\epsilon}\nonumber\\
&-i\sum_{n}\int_0^\infty\!\!dk\,g^2(k)\int_{-\infty}^{\infty}\!\frac{dp_0}{2\pi}
{\rm Tr}\left\{\sigma_n\frac{1}{p_0-(m-i\epsilon)\sigma_z}\right\}\sigma_n\frac{1}{-k^2+i\epsilon}
-\delta m\,\sigma_z,\\
&{\hat\Sigma}^{(2)}(p_0)
=i\int_0^\infty\!\!dk\,{\hat g}^2(k)\int_{-\infty}^{\infty}\!\frac{dk_0}{2\pi}
\sigma_x\frac{1}{p_0+k_0-(m-i\epsilon)\sigma_z}\sigma_x\frac{1}{k_0^2-k^2+i\epsilon}
-\delta{\hat m}\,\sigma_z,\\
&{\breve\Sigma}^{(2)}(p_0)
=i\sum_{n}\int_0^\infty\!\!dk\,{\hat g}^2(k)\int_{-\infty}^{\infty}\!\frac{dk_0}{2\pi}
\tau_n\left(\frac{{\mathbb P}_e}{p_0+k_0-m_e+i\epsilon}+\frac{{\mathbb
P}_g}{p_0+k_0-m_g-i\epsilon}\right)\tau_n\frac{1}{k_0^2-k^2+i\epsilon}
-\delta{\breve m},
\end{align}
\end{subequations}
where $\delta{\breve m}$ is the mass renormalization matrix with the eigenvalues $\delta m_e$ and $\delta m_g$.
With the use of the relations (\ref{algprop}) and (\ref{algprop1}) we can replace all matrices by the projectors
\begin{subequations}
\begin{align}\label{sigma2}
&\sum_{n}\sigma_n\frac{1}{p_0+k_0-(m-i\epsilon)\sigma_z}\sigma_n=\frac{2{\mathbb P}_e+{\mathbb P}_g}{p_0+k_0+m-i\epsilon}+\frac{2{\mathbb P}_g+{\mathbb P}_e}{p_0+k_0-m+i\epsilon},\\
&{\rm Tr}\left\{\sigma_n\frac{1}{p_0-(m-i\epsilon)\sigma_z}\right\}=\left\{\begin{array}{cc}
0&(n=x,y)\\
2m(p_0^2-4m^2+i\epsilon)^{-1}&(n=z)
\end{array}\right.,\\
&\sigma_x\frac{1}{p_0+k_0-(m-i\epsilon)\sigma_z}\sigma_x=\frac{{\mathbb P}_e}{p_0+k_0+m-i\epsilon}+\frac{{\mathbb P}_g}{p_0+k_0-m+i\epsilon},\\
&\sum_{n}\tau_n\left(\frac{{\mathbb P}_e}{p_0+k_0-m_e+i\epsilon}
+\frac{{\mathbb P}_g}{p_0+k_0-m_g-i\epsilon}\right)\tau_n
=\frac{{\mathbb P}_e}{p_0+k_0-m_g-i\epsilon}+\frac{3{\mathbb P}_g}{p_0+k_0-m_e+i\epsilon}
\end{align}
\end{subequations}
and then we can easily perform the integrations over $k_0$ ($m_\lambda$ will be equal either to $m_e$ or $m_g$)
\begin{subequations}
\begin{align}\label{sigma3}
i\int_{-\infty}^{\infty}\!\frac{dk_0}{2\pi}\frac{1}{p_0+k_0- m_\lambda\mp i\epsilon}\frac{1}{k_0^2-k^2+i\epsilon}=\frac{1}{2k(p_0\pm k- m_\lambda\mp i\epsilon)},\;\;
i\int_{-\infty}^{\infty}\!\frac{dp_0}{2\pi}\frac{2m}{p_0^2-4m^2+i\epsilon}=1.
\end{align}
\end{subequations}
Finally, we obtain
\begin{align}\label{fese}
\Sigma^{(2)}(p_0)=(2{\mathbb P}_e+{\mathbb P}_g)\int_0^\infty\!\!\frac{dk}{2k}\frac{g^2(k)}{p_0+k+m-i\epsilon}
+(2{\mathbb P}_g+{\mathbb P}_e)\int_0^\infty\!\!\frac{dk}{2k}\frac{g^2(k)}{p_0-k-m+i\epsilon}
+({\mathbb P}_e-{\mathbb P}_g)(m_t-\delta m),
\end{align}
\end{widetext}
where
\begin{align}\label{tad}
m_t=\int_0^\infty\!\!\frac{dk}{k^2} g^2(k).
\end{align}
Note that the contribution proportional to $m_t$, corresponding to the tadpole diagram, has the same form as the contribution from the mass correction.

For the two-level atom, we obtain
\begin{align}\label{fese1}
&{\hat\Sigma}^{(2)}(p_0)={\mathbb P}_e\int_0^\infty\!\!\frac{dk}{2k}\frac{\hat{g}^2(k)}{p_0+k+m-i\epsilon}\\
&-{\mathbb P}_g\int_0^\infty\!\!\frac{dk}{2k}\frac{\hat{g}^2(k)}{p_0-k-m+i\epsilon}
-({\mathbb P}_e-{\mathbb P}_g)\delta\hat{m}.
\end{align}

The electron self-energy part for the dipole atom is slightly more complicated
\begin{align}\label{fese2}
&{\breve\Sigma}^{(2)}(p_0)
={\mathbb P}_e\int_0^\infty\!\!\frac{dk}{2k}\frac{{\breve g}^2(k)}{p_0+k-m_g-i\epsilon}\\
&+3{\mathbb P}_g\int_0^\infty\!\!\frac{dk}{2k}\frac{{\breve g}^2(k)}{p_0-k-m_e+i\epsilon}
-{\mathbb P}_e\delta m_e-{\mathbb P}_g\delta m_g.
\end{align}

The mass corrections $\delta m$, $\delta{\hat m}$, and $\delta{\breve m}$ will be chosen so that the propagator $G(p_0)$ with radiative corrections has a pole at the renormalized mass. These pole conditions imply that $\Sigma^{(2)}(m\,\sigma_z)=0$ and ${\hat\Sigma}^{(2)}(m\,\sigma_z)=0$ and they give
\begin{subequations}
\begin{align}
\delta m&=\int_0^\infty\frac{dk}{2k^2}\,g^2(k)\frac{3k+2m}{k+2m},\\
\delta{\hat m}&=\int_0^\infty\frac{dk}{2k}\,{\hat g}^2(k)\frac{1}{k+2m}.
\end{align}
\end{subequations}
For the dipole atom the mass corrections are different for the ground state and for the excited state --- the energy of the excited state is raised and the energy of the ground state, as is always the case, is pushed down
\begin{subequations}\label{mcorr}
\begin{align}
\delta m_e&=\int_0^\infty\frac{dk}{2k}\,\frac{{\breve g}^2(k)}{k+\Delta m},\\
\delta m_g&=-3\int_0^\infty\frac{dk}{2k}\,\frac{{\breve g}^2(k)}{k+\Delta m},
\end{align}
\end{subequations}
where $\Delta m=m_e-m_g$. All these mass corrections give frequency-independent shifts in the level separation. The electron propagators do not have a direct physical interpretation but they serve as important ingredients in the calculation of the photon propagators. In particular, we will need the mass corrections to complete the calculation of the spin susceptibility and the atomic polarizability in the fourth order of perturbation theory.

\section{Photon propagator}\label{spp}

The photon propagator plays a distinguished role in our formulation, much more so than the electron propagator, since it enables one to calculate several important physical characteristics of two-level systems. The propagation of photons is, of course, modified by the presence of a two-level system. The scattering of photons off a two-level system is the counterpart of an important phenomenon in QED --- the vacuum polarization.

The relation of the full photon propagator to the self-energy part ${\Pi}_{ij}(k,k',k_0)$ is illustrated in Fig.~\ref{Fig5}b. It is slightly more complicated than in the case of the electron propagator because, in addition to a multiplication of matrices in the space of the vector components, we must perform an integration over the wave vector $k$. The counterpart of Eq.~(\ref{g}) is
\begin{align}\label{gp}
&{\cal G}_{Fij}(k,k',k_0)\nonumber\\
&=D_{Fij}(k,k',k_0)+\sum_{l}\int_0^\infty\!dk_1\sum_{n}\int_0^\infty\!dk_2\nonumber\\
&\times D_{Fil}(k,k_1,k_0){\Pi}_{ln}(k_1,k_2,k_0){\cal G}_{Fnj}(k_2,k',k_0).
\end{align}
Taking into account the fact that $D_{Fij}(k,k',k_0)$ is proportional to the Kronecker $\delta_{ij}$ and the Dirac $\delta(k-k')$, we can rewrite this equation in the form
\begin{align}\label{gp1}
&{\cal G}_{Fij}(k,k',k_0)
=\frac{\delta_{ij}\delta(k-k')}{k_0^2-k^2+i\epsilon}
+\frac{g(k)}{k_0^2-k^2+i\epsilon}\nonumber\\
&\times\sum_{l}\int_0^\infty\!dk''\,g(k'')
{\cal P}_{il}(k_0){\cal G}_{Flj}(k'',k',k_0),
\end{align}
where we took advantage of the factorization of ${\Pi}_{ij}(k,k',k_0)$
\begin{eqnarray}\label{pi2p}
{\Pi}_{ij}(k,k',k_0)=g(k){\cal P}_{ij}(k_0)g(k').
\end{eqnarray}
The iteration of Eq.~(\ref{gp1}) leads to the following expansion:
\begin{align}\label{gp2}
&{\cal G}_{Fij}(k,k',k_0)\nonumber\\
&=\frac{\delta_{ij}\delta(k-k')}{k_0^2-k^2+i\epsilon}
+\frac{g(k)}{k_0^2-k^2+i\epsilon}{\cal P}_{ij}(k_0)\frac{g(k')}{k_0^2-k'^{\,2}+i\epsilon}\nonumber\\
&+\frac{g(k)}{k_0^2-k^2+i\epsilon}{\cal P}_{il}(k_0)\int_0^\infty\!\!dk''\frac{g^2(k'')}{k_0^2-k''^{\,2}+i\epsilon}\nonumber\\
&\times{\cal P}_{lj}(k_0)\frac{g(k')}{k_0^2-k'^{\,2}+i\epsilon}+\dots.
\end{align}
This geometric series can be summed up and the final formula is
\begin{align}\label{gpf}
{\cal G}_{Fij}(k,k',k_0)&=D_{Fij}(k,k',k_0)\nonumber\\
&+\frac{g(k)}{k_0^2-k^2+i\epsilon}T_{ij}(k_0)\frac{g(k')}{k_0^2-k'^{\,2}+i\epsilon}.
\end{align}
The transition matrix $T(k_0)$ has the following representation in terms of the self-energy part:
\begin{align}\label{t}
T(k_0)=\frac{{\cal P}(k_0)}{1+{\cal P}(k_0)h(k_0)}
=\frac{1}{{\cal P}(k_0)^{-1}+h(k_0)},
\end{align}
where
\begin{align}\label{h}
h(k_0)=\int_0^\infty\!\frac{dk\,g^2(k)}{k^2-k_0^2-i\epsilon}.
\end{align}
Both $T(k_0)$ and ${\cal P}(k_0)$ in Eq.~(\ref{t}) are to be treated as $3\times 3$ matrices and the matrix to the power of $-1$ is meant as the inverse matrix.

The function $h(k_0)$ will play an important role in our calculations because in the lowest order of perturbation theory its real part determines the shift in the position of the resonance and the imaginary part determines the width of the resonance
\begin{align}\label{t2}
\mathrm{Re}\,h(k_0)&={\rm P}\int_0^\infty\!\frac{dk\,g^2(k)}{k^2-k_0^2},\\
\mathrm{Im}\,h(k_0)&=\frac{\pi g^2(|k_0|)}{2|k_0|}.
\end{align}
It follows from the assumptions that determine the validity of our model that the real part of $h(k_0)$ is practically constant and can be replaced by its value at 0 and the imaginary part varies as $k_0^3$. For example, when $\rho(r)$ and $g(k)$ are given by Eqs.~(\ref{gs}), we obtain
\begin{align}\label{gs1}
h(k_0)&=\frac{\mu^2\left(1+9\xi^2-9\xi^4-\xi^6+16i\xi^3\right)}{12\pi a_0^3(1+\xi^2)^4},
\end{align}
where $\xi=k_0a_0/2$. The value of the dimensionless parameter $\xi$ is very small in the range of wave vectors that cause the transitions between the two energy levels of our qubit. Thus, we can take only the leading terms and neglect higher powers of $\xi$ as compared to 1, to obtain
\begin{align}\label{h1}
h(k_0)\approx\frac{\mu^2}{12\pi a_0^3}+i\frac{\mu^2k_0^3}{6\pi}.
\end{align}
The formulas (\ref{gpf}) and (\ref{t}) are also valid for the two-level atom and the dipole atom. In both cases $h(k_0)$ is defined by Eq.~(\ref{h}) where $g(k)$ should be replaced either by ${\hat g}(k)$ or by ${\breve g}(k)$. Of course, in the first case there are no vector indices --- ${\hat T}(k_0)$ and ${\hat{\cal P}}(k_0)$ are not matrices but ordinary functions. In the second case, as is seen in Eq.~(\ref{pid1}) below, owing to the full rotational invariance, the matrix ${\breve{\cal P}}(k_0)$ is proportional to $\delta_{ij}$.

\subsection{Second order of perturbation theory}

\begin{figure}
\centering \vspace{0.5cm}
\includegraphics[scale=0.8]{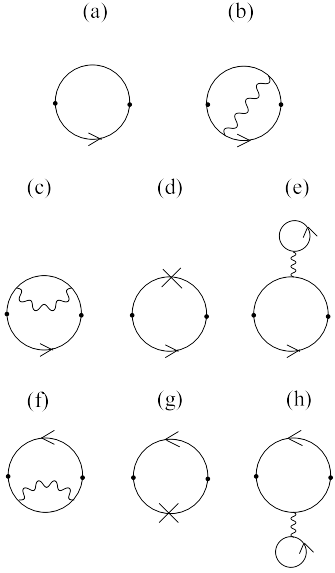}
\caption{Photon self-energy diagrams in the second and fourth order of perturbation theory.}\label{Fig6}
\end{figure}

In the second order, the radiative correction to the photon propagator is represented by the diagram (a) in Fig.~\ref{Fig6}. The photon self-energy part, constructed according to the rules given in Fig.~\ref{Fig3} has the form
\begin{align}\label{pi1}
{\cal P}_{ab}^{(2)}(k_0) = -i\int_{-\infty}^{\infty}\!\frac{dp_0}{2\pi}
\mathrm{Tr}\left\{\sigma_aS_F(p_0+k_0)\sigma_bS_F(p_0)\right\}.
\end{align}
The indices $(a,b)$ may take the values $x,y$ and $z$ in the Cartesian basis or the values $+,-$ and 0 in the angular momentum basis. The matrices ${\bm\sigma}$ are to be replaced by $\sigma_x$ for the two-level atom and by the matrices ${\bm\tau}$ in the case of the atomic dipole.

For the spin system it is convenient to choose the interaction Hamiltonian in the angular momentum basis (\ref{finham2}) because in this basis the photon self-energy part is diagonal. The components of the self-energy in the angular momentum basis are ${\cal P}_{\pm}(k_0)$ and ${\cal P}_{0}(k_0)$. They correspond to the following choices of the matrices $\sigma$ in Eq.~(\ref{pi1}):
\begin{subequations}\label{pi11}
\begin{align}
{\cal P}_{+}(k_0):\;\;\sigma_a=\sigma_-,\;\sigma_b=\sigma_+\\
{\cal P}_{-}(k_0):\;\;\sigma_a=\sigma_+,\;\sigma_b=\sigma_-\\
{\cal P}_{0}(k_0):\;\;\sigma_a=\sigma_z,\;\sigma_b=\sigma_z.
\end{align}
\end{subequations}
Making use of the properties (\ref{algprop}) of the $\sigma$ matrices, we end up with the following integrals:
\begin{subequations}\label{pi2}
\begin{align}
{\cal P}^{(2)}_{\pm}(k_0)
&=2\int_{-\infty}^{\infty}\!\frac{dp_0}{2\pi i}
\frac{1}{p_0+k_0\mp m\pm i\epsilon}\frac{1}{p_0\pm m\mp i\epsilon}\nonumber\\
&=-\frac{2}{2m\mp k_0},\\
{\cal P}^{(2)}_{0}(k_0)&=0.\label{pi2b}
\end{align}
\end{subequations}
The component ${\cal P}^{(2)}_{0}(k_0)$ vanishes because in the corresponding integrals both residues lie in the same half-plane. The relation ${\cal P}^{(2)}_{-}(k_0)={\cal P}^{(2)}_{+}(-k_0)$ is a direct confirmation of the time-reversal invariance. The angular momentum components of the transition matrix $T(k_0)$ obtained by substituting these self-energy parts into Eq.~(\ref{t}) are
\begin{subequations}\label{ts}
\begin{align}
T_{\pm}^{(2)}(k_0)&=-\frac{2}{2m\mp k_0-2h(k_0)}\\
T_{0}^{(2)}(k_0)&=0.
\end{align}
\end{subequations}

For the two-level atom we must take $a=x$ and $b=x$ in Eq.~(\ref{pi1}). After evaluating the trace, the integral reduces to the sum of two simple integrals
\begin{align}\label{pse2}
&{\hat{\cal P}}^{(2)}(k_0) = -i\int_{-\infty}^{\infty}\!\frac{dp_0}{2\pi}\nonumber\\
&\times\mathrm{Tr}\left\{\sigma_x\frac{1}{p_0+k_0-(m-i\epsilon)\sigma_z}\sigma_x
\frac{1}{p_0-(m-i\epsilon)\sigma_z}\right\}\nonumber\\
&=\int_{-\infty}^{\infty}\!\frac{dp_0}{2\pi i}
\frac{1}{p_0+k_0+m-i\epsilon}\,\frac{1}{p_0-m+i\epsilon}\nonumber\\
&+\int_{-\infty}^{\infty}\!\frac{dp_0}{2\pi i}\frac{1}{p_0+k_0-m+i\epsilon}\,\frac{1}{p_0+m-i\epsilon}.
\end{align}
The result of the integrations is
\begin{align}\label{pse3}
{\hat{\cal P}}^{(2)}(k_0)=-\frac{4m}{4m^2-k_0^2}
\end{align}
and it leads to the following formula for $\hat{T}(k_0)$ in the lowest order of perturbation theory:
\begin{align}\label{ta1}
\hat{T}^{(2)}(k_0)=-\frac{4m}{4m^2-k_0^2-4m{\hat h}(k_0)}.
\end{align}

For the dipole atom the contribution represented by the diagram (a) in Fig.~\ref{Fig6} leads to the following expression for the self-energy part:
\begin{align}\label{pid1}
{\breve{\cal P}}_{ij}^{(2)}(k_0)&=\delta_{ij}\int_{-\infty}^{\infty}\!\frac{dp_0}{2\pi i}
\frac{1}{p_0+k_0-m_e+i\epsilon}\frac{1}{p_0-m_g-i\epsilon}\nonumber\\
&+\delta_{ij}\int_{-\infty}^{\infty}\!\frac{dp_0}{2\pi i}
\frac{1}{p_0+k_0-m_g-i\epsilon}\frac{1}{p_0-m_e+i\epsilon}\nonumber\\
&=-\delta_{ij}\frac{2\Delta m}{\Delta m^2-k_0^2}.
\end{align}
This leads to the transition matrix of the form
\begin{align}\label{ta}
\breve{T}^{(2)}_{ij}(k_0)=-\delta_{ij}\frac{2\Delta m}{\Delta m^2-k_0^2-2\Delta m{\breve h}(k_0)}.
\end{align}

\subsection{Fourth order of perturbation theory}

The calculation of the photon self-energy part up to the fourth-order of perturbation theory requires the evaluation of all the contributions to the photon self-energy represented by the Feynman diagrams (b)--(h) shown in Fig.~\ref{Fig6}. These calculations are presented in Appendix \ref{a2}. Upon substituting the results of this calculation into Eq.~(\ref{t}), we obtain the formula for the transition matrix. However, there is an additional problem now that has not been present in the calculation of the self-energy part in the lowest order. In the formulas for the self-energy parts (\ref{all1}) of the spin system we encounter double poles $1/(2m-k_0)^2$ and $1/(2m+k_0)^2$. Such terms indicate a breakdown of the simple perturbation theory since for $k_0\approx 2m$ fourth-order terms dominate over second-order terms. The remedy comes from the realization that these double poles simply indicate an additional shift in the position of the resonance. Indeed, expanding $1/(2m-k_0-\delta)$ into powers of $\delta$ we encounter higher-order poles. We encountered the same problem in the expansion of the electron and photon propagators into the perturbation series but the difference is that then we were able to sum up the whole geometric series. Here, we can do it only order by order. In the present case, we can eliminate the double pole in the formulas (\ref{all}) by the following substitution:
\begin{align}\label{rear}
\frac{1}{2m-k_0}+\frac{\delta}{(2m-k_0)^2} \to \frac{1}{2m-k_0-\delta},
\end{align}
that reproduces correctly the lowest order correction in $\delta$. Higher powers of $\delta$ contribute to higher orders of perturbation theory. Applying this procedure to the expressions for the self-energy parts (\ref{all}) we obtain the following formulas that do not suffer from the double-pole problem:
\begin{subequations}\label{all5}
\begin{align}
&{\cal P}_{\pm}^{(2+4)}(k_0)=-\frac{2(1-b)}{2m\mp k_0-\delta},\\
&{\cal P}_{0}^{(2+4)}(k_0)={\cal P}_{0}^{(4)}(k_0)\nonumber\\
&=-4\int_0^\infty\!\frac{dk}{k}\frac{g^2(k)}{k+2m}\,\frac{1}{(k+2m)^2-k_0^2-i\epsilon}.
\end{align}
\end{subequations}
Therefore, the transition matrix in this order is
\begin{subequations}\label{t5}
\begin{align}
&{T}_{\pm}^{(2+4)}(k_0)=-\frac{2(1-b)}{2m\mp k_0-\delta-2(1-b)h(k_0)},\label{t5a}\\
&{T}_{0}^{(2+4)}(k_0)={\cal P}_{0}^{(2+4)}(k_0).\label{t5b}
\end{align}
\end{subequations}
The last equation follows from the fact that, as seen from Eq.~(\ref{pi2b}), ${\cal P}_{0}$ does not contain terms of the second order.

The results for the two-level atom in the fourth order are even simpler since there is only one component of the self-energy part and there are no double poles. Substituting the self-energy part (\ref{pse2l}) into the formula (\ref{t}), we obtain
\begin{align}\label{f5}
\hat{T}^{(2+4)}(k_0)=-\frac{4m(1-\hat{b})}{4m^2-k_0^2-4m(1-\hat{b})\hat{h}(k_0)}.
\end{align}

The transition matrix for the dipole atom in the fourth-order, obtained from the self-energy part (\ref{pseda}), has the same general form as that for the two-level atom
\begin{align}\label{fb5}
\breve{T}^{(2+4)}_{ij}(k_0)=-\delta_{ij}\frac{2\Delta m(1-\breve{b})}{\Delta m^2-k_0^2-2\Delta m(1-\breve{b})\breve{h}(k_0)}.
\end{align}

\section{Photon scattering amplitude}

The photon scattering amplitude $f_{ij}(\omega)$ can be obtained \cite{bd,iz} from the photon propagator (\ref{gpf}) by stripping off the free propagators at both ends and putting the whole expression on the energy shell $k_0=\omega,\,k= \omega,\,k'=\omega$. Therefore the scattering amplitude is related to the transition matrix by the formula
\begin{align}\label{scatta}
f_{ij}(\omega)=g^2(\omega)T_{ij}(\omega).
\end{align}
The argument $\omega$ of the scattering amplitude can only take positive values because the photon energy is positive.

In the second order of perturbation theory, the self-energy part for the spin system is given by Eq.~(\ref{pi2}). Therefore, according to Eq.~(\ref{scatta}), the photon scattering amplitude for a spin system is
\begin{subequations}\label{t4}
\begin{align}
f_\pm^{(2)}(\omega)&=-\frac{2g^2(\omega)}{2m-\Delta(\omega)\mp\omega-i\Gamma(\omega)},\\
f_0^{(2)}(\omega)&=0,
\end{align}
\end{subequations}
where the energy-dependent shift and width in this order according to Eqs.~(\ref{t2}) are
\begin{align}\label{ws}
{\Delta}(\omega)&={\rm P}\int_0^\infty\!\frac{dk\,2g^2(k)}{k^2-\omega^2},\\
{\Gamma}(\omega)&=\frac{\pi g^2(\omega)}{|\omega|}.
\end{align}
Owing to the angular momentum conservation, the photons with $M_z=0$ do not scatter in the lowest order because such photons cannot cause a direct transition from the lower to the upper state. They may cause such a transition provided it is accompanied by a simultaneous emission of a photon with $M_z=-1$ but this is a higher order process. Indeed, in the fourth order the scattering amplitude for $M_z=0$, as seen from Eq.~({\ref{t5b}), does not vanish. In comparison with the standard Breit-Wigner resonance formula, our $\Gamma(\omega)$ is equal to half-width. As a result of the angular momentum conservation, the amplitude with $M_z=1$ is resonant but the one with $M_z=-1$ is not. These two amplitudes correspond to two possible time orderings illustrated in Fig.~\ref{Fig2}.

The amplitude for the scattering of a photon off a two-level atom in the second order is obtained from Eq.~(\ref{pse3}) for the transition matrix
\begin{align}\label{t1}
{\hat f}^{(2)}(\omega)=-\frac{4m{\hat g}^2(\omega)}{4m^2-\omega^2-4m{\hat\Delta}(\omega)-4mi{\hat\Gamma}(\omega)},
\end{align}
where
\begin{align}\label{ws1}
{\hat\Delta}(\omega)&={\rm P}\int_0^\infty\!\frac{dk\,\hat{g}^2(k)}{k^2-\omega^2},\\
{\hat\Gamma}(\omega)&=\frac{\pi \hat{g}^2(\omega)}{2|\omega|}.
\end{align}
The standard resonance behavior can be seen after rewriting ${\hat f}(\omega)$ in a different form. Disregarding the square of ${\hat\Delta}(\omega)+i{\hat\Gamma}(\omega)$ and its product with $\hat{g}^2$ since they are both of the fourth-order, we can approximately decompose the scattering amplitude (\ref{t1}) into the following sum of simple fractions:
\begin{align}\label{t3}
{\hat f}^{(2)}(\omega)\approx-\frac{\hat{g}^2(\omega)}{2m-{\hat\Delta}(\omega)-\omega-i{\hat\Gamma}(\omega)}\nonumber\\
-\frac{\hat{g}^2(\omega)}{2m-{\hat\Delta}(\omega)+\omega-i{\hat\Gamma}(\omega)}.
\end{align}
The first term is clearly resonant and the second is not. This expression agrees with the equal sign prescription --- the width ${\hat\Gamma}(\omega)$ enters with {\em the same sign} in both terms. Thus, the equal sign prescription, advocated in Ref.~\cite{sted2}, is appropriate for the photon scattering amplitude but not for the polarizability, as will be explained later. The scattering amplitude off a dipole atom has the same general form as for the two-level atom, so that our analysis applies also to this case.

\section{Linear response theory}

We shall use the quantum linear response theory (cf., for example, \cite{fw}) to relate the Feynman propagators to the important physical characteristics of two-level systems: the spin susceptibility and the atomic polarizability. Linear response theory describes the reaction of a quantum system to a weak external perturbation. In the linear response theory, changes in an expectation values of observables are expressed in terms of retarded propagators. In our opinion most of the controversies in the treatment of damping resulted from the lack of a clear distinction between the scattering of photons (described by the S matrix and the Feynman propagators) and the time evolution of the expectation values (described by the solutions of the Heisenberg equations of motion and the retarded propagators). For the spin system, the observables are the spin (or the magnetic moment) components $\Psi^\dagger{\bm\sigma}\Psi$. For the two-level atom, the observable is the atomic induced dipole represented (up to a constant factor) by the operator $\Psi^\dagger{\sigma}_x\Psi$. The spin susceptibility determines the response of the magnetic moment to the applied magnetic field and the polarizability determines the response of the atom to the applied electromagnetic field. These external perturbations are represented by the terms  $\Psi^\dagger{\bm\sigma}\Psi\!\cdot\!\delta{\bm\varphi}(t)$ or $\Psi^\dagger\sigma_x\Psi\delta{\varphi}(t)$ added to the Hamiltonian. The change in the average spin value produced by such a perturbation is \cite{fw}
\begin{align}\label{change}
&\delta\langle G|S_i(t)|G\rangle\nonumber\\
&=-i\int_{-\infty}^{\infty}\!dt'\theta(t-t')\langle G|[S_i(t),S_j(t')]|G\rangle \delta\varphi_j(t').
\end{align}
The spin operators in this formula are in the Heisenberg picture ${\bm S}(t)=\Psi^\dagger(t){\bm\sigma}\Psi(t)$. Since the ground state $|G\rangle$ is stationary with some energy $E_G$, we can write
\begin{align}\label{stat}
&\langle G|[S_i(t),S_j(t')]|G\rangle\nonumber\\
&=\langle G|S_i(0)e^{-i(H-E_G)(t-t')}S_j(0)|G\rangle\nonumber\\
&-\langle G|S_j(0)e^{i(H-E_G)(t-t')}S_i(0)|G\rangle.
\end{align}
Thus, the integral in Eq.~(\ref{change}) is a convolution and we can transform this equation to a simple algebraic relation between the Fourier transforms
\begin{align}\label{rft}
\delta\langle G|{\tilde S}_i(\omega)|G\rangle =\chi_{ij}(\omega)\delta{\tilde\varphi}_j(\omega).
\end{align}
The spin susceptibility $\chi_{ij}(\omega)$ describes the response of the spin to a monochromatic external magnetic field and is defined by the Kubo formula \cite{kubo}
\begin{align}\label{susc}
\chi_{ij}(\omega)=-i\int_{-\infty}^0\!dt\,e^{-i\omega t+\epsilon t}\langle G|[S_i(0),S_j(t)]|G\rangle,
\end{align}
where the damping factor $e^{\epsilon t}$ guarantees that the applied field has been switched on adiabatically.

The corresponding formula for the polarizability of a two-level atom reads
\begin{align}
&\delta\langle G|S_x(t)|G\rangle\nonumber\\
&=-i\int_{-\infty}^{\infty}\!dt'\theta(t-t')\langle G|[S_x(t),S_x(t')]|G\rangle \delta\varphi(t'),\label{change1}\\
&\alpha(\omega)=iA\int_{-\infty}^0\!dt e^{-i\omega t+\epsilon t}\langle G|[S_x(0),S_x(t)]|G\rangle.\label{change2}
\end{align}
For a single two-level atom, the constant factor $A$ is usually (cf., for example, \cite{mb,bbm}) given the value $A=d^2/3\hslash$, where $d$ measures the strength of the dipole transition. We have reversed the sign in the definition of $\alpha(\omega)$, as compared to the definition of the spin susceptibility (\ref{susc}), to be in agreement with the standard Kramers-Heisenberg-Dirac expression for the polarizability.

The expectation value of the retarded commutator of the spin operators appearing in (\ref{change}) is directly related to the retarded photon propagator ${\cal G}_{Rij}$
\begin{align}\label{retph}
{\cal G}_{Rij}(k,k',t-t')=-i\theta(t-t')\langle G|[\Phi_i(t),\Phi_j(t')]|G\rangle.
\end{align}
Indeed, with the use of the Heisenberg equations of motion (\ref{heqm2}) for the field ${\bm\Phi}$ and the canonical commutation relations (\ref{ccr2}), we obtain
\begin{align}\label{rel}
&(\partial_t^2+k^2)(\partial_t'^{\,2}+k'^{\,2}){\cal G}_{Rij}(k,k',t-t')\nonumber\\
&=-(\partial_t^2+k^2)\delta(t-t')\delta_{ij}\delta(k-k')\nonumber\\
&-ig(k)g(k')\theta(t-t')\langle G|[S_i(t),S_j(t')]|G\rangle.
\end{align}
After the Fourier transformation with respect to $t-t'$, this relation becomes
\begin{align}\label{frel}
&(k_0^2-k^2)(k_0^2-k'^{\,2}){\cal G}_{Rij}(k,k',\omega)\nonumber\\
&=(k_0^2-k^2)\delta_{ij}\delta(k-k')+g(k)g(k')\chi_{ij}(\omega).
\end{align}
Analogous relations hold for the two-level atom. Thus, the susceptibility and the polarizability are simply proportional to the retarded photon propagator.

\subsection{Spectral representation}

One the advantages of using the methods of relativistic field theory is that the analytic properties of propagators became explicit. The retarded photon propagator {\em cannot be calculated} by a direct application of the Feynman-Dyson perturbation theory. However, once we determine the photon Feynman propagator, the retarded propagator can be unambiguously reconstructed. In order to prove this assertion, we shall follow the same procedure that in a relativistic quantum field theory leads to the K{\"a}llen-Lehmann representation. Starting from the general definition of the photon Feynman propagator we arrive at the following formula (cf., for example, \cite{fw}):
\begin{align}\label{ppros}
{\cal G}_{Fij}&(k,k',t-t')=-i\theta(t-t')\langle G|\Phi_i(k,t)\Phi_j(k',t')|G\rangle\nonumber\\
&-i\theta(t'-t)\langle G|\Phi_i(k',t')\Phi_j(k,t)|G\rangle\nonumber\\
&=-i\theta(t-t')\int_0^\infty\!\!dM\,e^{-iM(t-t')}A_{ij}(M,k,k')\nonumber\\
&-i\theta(t'-t)\int_0^\infty\!\!dM\,e^{-iM(t-t')}A_{ji}(M,k',k),
\end{align}
where the spectral matrix $A_{ij}(M,k,k')$ is defined as follows:
\begin{align}\label{spec}
&A_{ij}(M,k,k')=A^*_{ji}(M,k',k)\nonumber\\
&=\sum_n\delta(M-E_n+E_G)\langle G|\Phi_i(k,0)|n\rangle \langle n|\Phi_j(k',0)|G\rangle
\end{align}
$\{|n\rangle\}$ is any complete set of stationary states of the system and $E_G$ is the energy of the ground state. Therefore, the Fourier transform ${\cal G}_{Fij}(k,k',k_0)$ of the propagator can be written in the form of a spectral representation
\begin{align}\label{ppros1}
&{\cal G}_{Fij}(k,k',k_0)\nonumber\\
&=\int_0^\infty\!dM\,\left(\frac{A_{ij}(M,k,k')}{k_0-M+i\epsilon}
-\frac{A^*_{ij}(M,k,k')}{k_0+M-i\epsilon}\right).
\end{align}
Repeating the same procedure for the retarded propagator defined in Eq.~(\ref{retph}), we obtain its spectral representation
\begin{align}\label{retprs}
&{\cal G}_{Rij}(k,k',k_0)\nonumber\\
&=\int_0^\infty\!dM\,\left(\frac{A_{ij}(M,k,k')}{k_0-M+i\epsilon}
-\frac{A^*_{ij}(M,k,k')}{k_0+M+i\epsilon}\right).
\end{align}
The spectral matrices are the same and {\em the only difference} is the change of the sign of the $\epsilon$-term in the second part. Since $M\ge 0$, this is equivalent to the replacement of $i\epsilon$ in the denominators of the Feynman propagator by $i{\rm sgn}(k_0)\epsilon$. This ``epsilon rule'' in the simplest case of a one-component propagator reduces to the standard rule of quantum linear response theory \cite{fw}:
\begin{subequations}\label{rvsf}
\begin{align}
{\rm Re}\,{\cal G}_{R}(k,k',k_0)&={\rm Re}\,{\cal G}_{F}(k,k',k_0),\\
{\rm Im}\,{\cal G}_{R}(k,k',k_0)&={\rm sgn}(k_0){\rm Im}\,{\cal G}_{F}(k,k',k_0).
\end{align}
\end{subequations}
These simple relations do not hold in general because the imaginary unit may appear not only together with $\epsilon$ but also in other places. In particular, they do not hold for the spin system in the Cartesian basis. However, in the angular momentum basis the photon propagator is diagonal. Therefore, we can treat each component separately and use the relations (\ref{rvsf}) to obtain the components of susceptibility from the Feynman propagator (or more precisely from the scattering matrix)
\begin{subequations}\label{susc1}
\begin{align}
{\rm Re}\,\chi_{a}(\omega)&={\rm Re}\,T_{a}(\omega),\\
{\rm Im}\,\chi_{a}(\omega)&={\rm sgn}(\omega){\rm Im}\,T_{a}(\omega),
\end{align}
\end{subequations}
where the subscript $a$ takes on the values $\pm$ and $0$. Here, unlike in the formula (\ref{scatta}) for the scattering amplitude, in Eqs.~(\ref{susc1}) the frequency $\omega$ takes on positive and negative values because the real function $\varphi(t)$ describing an external perturbation must contain both positive and negative frequencies. This important difference must be kept in mind when discussing the relations between the scattering and the linear response.

The relations between the Feynman propagator and the retarded propagator in the simple form (\ref{rvsf}) hold for the two-level atom and for the dipole atom. In the first cases there is just one scalar function from the very beginning. In the second case, due to the conservation of the three components of angular momentum, the propagator is proportional to $\delta_{ij}$ so that it effectively reduces to just one function.

We shall confirm the validity of the spectral representations of the photon propagator by direct calculations in perturbation theory.

Since our theory is invariant under time reversal, the retarded propagator will automatically satisfy the general crossing relation
\begin{align}\label{cross}
{\cal G}_{Rij}(k,k',-k_0)&={\cal G}^*_{Rji}(k,k',k_0).
\end{align}
This relation for a two-level atom implies that the atomic polarizability satisfied the condition $\alpha^*(\omega)=\alpha(-\omega)$. The significance of this condition has been emphasized in Ref.~\cite{lb}.

\subsection{Spin susceptibility}\label{ss}

The components of the photon transition matrix $T(k_0)$ evaluated in Sec.~\ref{spp} gives the following formula for the spin susceptibility in the lowest order of perturbation theory:
\begin{subequations}\label{chi2}
\begin{align}
&\chi_\pm^{(2)}(\omega)=-\frac{2}{2m\mp\omega-2\left(\Delta(\omega)\pm i\,{\rm sgn}(\omega)\Gamma(\omega)\right)},\\
&\chi_0^{(2)}(\omega)=0.
\end{align}
\end{subequations}
From the transition matrix (\ref{t5}) we can obtain the spin susceptibility in the fourth order
\begin{subequations}\label{chi4}
\begin{align}
&\chi_\pm^{(2+4)}(\omega)\nonumber\\
&=-\frac{2(1-b)}{2m\mp\omega-\delta-2(1-b)\left(\Delta(\omega)\pm i\,{\rm sgn}(\omega)\Gamma(\omega)\right)},\\
&\chi_0^{(2+4)}(\omega)={\cal P}_0^{(2+4)}(\omega).
\end{align}
\end{subequations}
There is only one term for each transition but the opposite sign prescription is still visible. The sign of the imaginary part in the denominator depends on the sign of $\omega$.

\subsection{Atomic polarizability}\label{ap}

In this Section we shall use the Hamiltonian (\ref{hamlb}) for the two-level atom to calculate the photon propagator up to the fourth order of perturbation theory and use it to find the polarizability. All Feynman diagrams corresponding to the radiative corrections that will be taken into account in our calculation are shown in Fig.~\ref{Fig6}. The frequency-dependent atomic polarizability $\alpha(\omega)$ can be obtained from the functions ${\hat T}(k_0)$ and  ${\breve T}(k_0)$ by changing their imaginary parts according to the prescription (\ref{susc1}).

The second-order self-energy part for the two-level atom is given in Eq.~(\ref{pse3}). Substituting this expression into the formula (\ref{t}) and changing the sign of the imaginary part, like in Eq.~(\ref{susc1}), we obtain the following expression for the atomic polarizability (\ref{change2}):
\begin{align}\label{pol1}
&\alpha^{(2)}(\omega)\nonumber\\
&=\frac{4mA}{4m^2-\omega^2-4m\left({\hat\Delta}(\omega)+i\,{\rm sgn}(\omega){\hat\Gamma}(\omega)\right)}.
\end{align}
In order to see that this expression obeys the opposite sign prescription we could write it in the form of a spectral representation. However, following the treatment of the photon scattering amplitude, we shall convert this expression into simple fractions neglecting again higher-order terms
\begin{align}\label{pol2}
&\alpha^{(2)}(\omega)\approx\frac{A}{2m-{\hat\Delta}(\omega)-\omega-i\,{\rm sgn}(\omega){\hat\Gamma}(\omega)}\nonumber\\
&+\frac{A}{2m-{\hat\Delta}(\omega)+\omega-i\,{\rm sgn}(\omega){\hat\Gamma}(\omega)}.
\end{align}
Depending on the sign of $\omega$, either the first or the second term is resonant. Therefore, if we only care about the important resonant terms, we can write this formula as
\begin{align}\label{pol2a}
&\alpha^{(2)}(\omega)\approx\frac{A}{2m-{\hat\Delta}(\omega)-\omega-i{\hat\Gamma}(\omega)}\nonumber\\
&+\frac{A}{2m-{\hat\Delta}(\omega)+\omega+i{\hat\Gamma}(\omega)}.
\end{align}
In other words, this expression is a good approximation to the exact formula (\ref{pol1}) near both resonances, when $\omega\approx\pm 2m$. Thus, our expressions for the atomic polarizability derived from the quantum linear response theory agree with the {\em opposite sign prescription}, as advocated in Refs.~\cite{bf,mb,bbm}.

To extend this result to the fourth order of perturbation theory, we use the formula (\ref{f5}) for the photon propagator. The resulting expression differs from the formula (\ref{pol1}) for the polarizability in the second order {\em only} by the presence of the factors $(1-{\hat b})$
\begin{align}\label{pol3}
&\alpha^{(2+4)}(\omega)\nonumber\\
&=\frac{4m(1-{\hat b})A}{4m^2-\omega^2-4m(1-{\hat b})\left({\hat\Delta}(\omega)+i\,{\rm sgn}(\omega){\hat\Gamma}(\omega)\right)}.
\end{align}
Our result is quite different from the formula derived in Ref.~\cite{lb}. It seems to us that this difference is due to the difficulties in systematically accounting in the standard treatment for all higher order corrections. In particular, Loudon and Barnett have not included all corrections to the ground state up to the fourth order and they have disregarded all level shifts. In our formulation, the method of Feynman diagrams guarantees an unambiguous derivation of all corrections in any order of perturbation theory.

Finally, we would like to emphasize that all those equal or opposite sign prescriptions, that are widely used in the semiphenomenological treatment, have some practical limitations. They can be directly applied only to the spectral representations (\ref{ppros1}) or (\ref{retprs}). In general, as seen for example in Eq.~(\ref{pol3}), the expressions obtained directly in perturbation theory cannot be easily decomposed into two parts because they are not given in the form of a spectral representation. Of course, we can always find this representation, but the formulas are quite complicated and they hide the resonant character of the process. Still, having closed expressions we can always identify the analytic properties of $\alpha(\omega)$ that correspond to these prescriptions. Namely, we can locate the positions of the poles of $\alpha(\omega)$ in the complex $\omega$ plane to discover that the pole near $2m$ lies in the {\em upper} half-plane while the the pole near $-2m$ lies in {\em lower} half-plane. This property extends the opposite sign prescription to the general case.

\section{Conclusions}

We have shown that the methods of relativistic quantum field theory applied to a two-level (and also to a many-level) system interacting with the quantized electromagnetic field lead to significant simplifications in the evaluation of various physical properties of the system. Owing to these simplifications we can easily go easily beyond the lowest orders of perturbation theory. The difference in complexity of the calculations performed with the use of the traditional approach and the new methods is enormous. For example, the interaction Hamiltonian for the spin system has six terms, so there are $6^4=1296$ terms in the fourth order of the standard perturbation theory while in our approach we have only several Feynman diagrams to consider. It is true that for a particular process many terms will not contribute but still a lot of terms must be taken into account. In addition to the simplifications in the calculations, we also gain new physical insights that stem from the connections that exist in quantum field theory between different characteristics of the system. In particular, the connection between the photon scattering amplitude and the linear response functions of the two-level system to an applied electromagnetic field is very useful. This connection is crucial to the understanding of the hotly debated relation between the equal sign prescription and the opposite sign prescription in the description of damping.

\acknowledgments

We acknowledge the support by the Polish Ministry of Science and Higher Education under the Grant for Quantum Information and Quantum Engineering.

\appendix

\section{Multipole expansion}\label{a1}

\subsection{Multipole expansion of the electromagnetic field}

The decomposition of the electric and magnetic field operators into the eigenfunctions of the angular momentum can be written in the form \cite{qed,jack}
\begin{align}
{\bm E}({\bm r})=\!\sum_{JM\lambda}\!\int_0^\infty\!\!\!dk\!\left[{\bm E}_{JMk}^{(\lambda)}({\bm r})c_{JM}^{(\lambda)}(k)+{\bm E}_{JMk}^{*(\lambda)}({\bm r})c_{JM}^{\dagger(\lambda)}(k)\right]\!,\label{dece}\\
{\bm B}({\bm r})=\!\sum_{JM\lambda}\!\int_0^\infty\!\!\!dk\!\left[{\bm B}_{JMk}^{(\lambda)}({\bm r})c_{JM}^{(\lambda)}(k)+{\bm B}_{JMk}^{*(\lambda)}({\bm r})c_{JM}^{\dagger(\lambda)}(k)\right]\!,\label{decm}
\end{align}
where $J>0$ and $\lambda$ takes on two values $(e,m)$ that distinguish between the electric and magnetic multipoles. The operators $c_{JMk}^{(\lambda)}(k)$ and $c_{JMk}^{\dagger(\lambda)}(k)$ annihilate and create photons with the energy $\hslash\omega=\hslash ck$, the square of the total angular momentum $\hslash^2J(J+1)$, the projection of the total angular momentum on the $z$ axis $\hslash M$, and the parity as determined by $\lambda$. These operators obey the standard commutation relations
\begin{eqnarray}\label{cr}
\left[c_{JM}^{(\lambda)}(k), c_{J'M'}^{\dagger(\lambda')}(k')\right] =\delta_{JJ'}\delta_{MM'}\delta_{\lambda\lambda'}\delta(k-k').
\end{eqnarray}
The energy operator of the electromagnetic field expressed in terms of the creation and annihilation operators is
\begin{eqnarray}\label{en}
H_f=\sum_{JM\lambda}\int_0^\infty\!dk\,\hslash\omega\, c_{JM}^{\dagger(\lambda)}(k)c_{JM}^{(\lambda)}(k).
\end{eqnarray}
The functions appearing in this decomposition (\ref{dece}) and (\ref{decm}) can be expressed in terms of the following solutions $T_{JMk}({\bm r})$ of the scalar Helmholtz equation:
\begin{eqnarray}\label{eigen}
T_{JMk}({\bm r})=\sqrt{\frac{k}{\pi J(J+1)}}j_J(k r)Y_{JM}({\bm n}),
\end{eqnarray}
where $j_J(k r)$ is the spherical Bessel function, $Y_{JM}({\bm n})$ is the spherical harmonic, and ${\bm n}$ is the unit vector in the ${\bm r}$ direction. The functions ${\bm B}_{JMk}^{(\lambda)}({\bm r})$ and ${\bm E}_{JMk}^{(\lambda)}({\bm r})$ can be expressed in terms of ${\bm T}_{JMk}({\bm r})$ as follows \cite{qed,jack}:
\begin{subequations}
\begin{eqnarray}\label{elmag}
{\bm E}_{JMk}^{(e)}({\bm r})&=&i{\bm\nabla}{\bm\times}{\bm L}\,T_{JMk}({\bm r}),\label{eel1}\\
{\bm B}_{JMk}^{(e)}({\bm r})&=&k{\bm L}\,T_{JMk}({\bm r}),\label{bel1}\\
{\bm E}_{JMk}^{(m)}({\bm r})&=&k{\bm L}\,T_{JMk}({\bm r}),\label{emag1}\\
{\bm B}_{JMk}^{(m)}({\bm r})&=&-i{\bm\nabla}{\bm\times}{\bm L}\,T_{JMk}({\bm r}),\label{bmag1}
\end{eqnarray}
\end{subequations}
where ${\bm L}=-i{\bm r}\times{\bm \nabla}$ is the angular momentum operator.

\subsection{Magnetic dipole coupling}

In order to extract the relevant part of ${\bm B}({\bm r})$ that will contribute to the interaction Hamiltonian, we shall insert the expansion (\ref{decm}) into the formula (\ref{ham1}) and obtain the following integrals (and their complex conjugate counterparts):
\begin{subequations}\label{md}
\begin{align}
\int\!d^3r\,\rho(r){\bm B}_{JMk}^{(e)}({\bm r}) &=\int\!d^3r\,\rho(r)k{\bm L}\,T_{JMk}({\bm r}),\label{bel}\\
\int\!d^3r\,\rho(r){\bm B}_{JMk}^{(m)}({\bm r}) &=-i\!\int\!d^3r\,\rho(r){\bm\nabla}{\bm\times}{\bm L}\,T_{JMk}({\bm r}).\label{bmag}
\end{align}
\end{subequations}
The integral in Eq.~(\ref{bel}) vanishes because after the integration by parts the angular momentum operator ${\bm L}$ acts on the spherically symmetric function $\rho(r)$. However, the integral in Eq.~(\ref{bmag}) might contribute. In order to explore this possibility, we use the identity $i{\bm\nabla}{\bm\times}{\bm L}=-i{\bm L}{\bm\times}{\bm\nabla}-2{\bm\nabla}$. Again, after the integration by parts we find that the term $i{\bm L}{\bm\times}{\bm\nabla}$ does not contribute. We are left only with the gradient term ${\bm\nabla}\rho(r)=\rho'(r){\bm n}$ and we obtain
\begin{align}\label{avb}
\int\!d^3r\,\rho(r){\bm B}_{JMk}^{(m)}({\bm r}) =2\int\!d^3r\,\rho'(r){\bm n}\,T_{JMk}({\bm r}).
\end{align}
The integral over the angles will give a nonvanishing contribution only when the spherical harmonic $Y_{JM}({\bm n})$ is a linear combination of the components of the vector ${\bm n}$, i.e. for $J=1$. Therefore, only the magnetic dipole component of the magnetic field contributes to the interaction Hamiltonian (\ref{ham1}). Using Eq.~(\ref{avb}) we can rewrite the interaction Hamiltonian (\ref{ham1}) in the form
\begin{align}\label{avb1}
H_I&=-\mu{\bm\psi}^\dagger{\bm\sigma}{\bm\psi}\!\cdot\!\int\!d^3r\,\sum_M\int_0^{\infty}
dk\,\sqrt{\frac{2k}{\pi}}j_1(kr)\rho'(r)\nonumber\\
&\times{\bm n}\left[Y_{1M}({\bm n})c_M(k)+Y_{1M}^*({\bm n})c_M^\dagger(k)\right].
\end{align}
To simplify the formulas we introduced the following notation:
\begin{align}\label{c}
c_+(k)=c_{1,1}^{(m)}(k),\;\;\;c_0(k)=c_{1,0}^{(m)}(k),\;\;\;c_-(k)=c_{1,-1}^{(m)}(k).
\end{align}
Using the explicit expressions for the spherical harmonics
\begin{subequations}
\begin{eqnarray}\label{spharm}
Y_{J=1,1}({\bm n})&=&\sqrt{\frac{3}{8\pi}}\frac{-x-iy}{r},\\
Y_{J=1,-1}({\bm n})&=&\sqrt{\frac{3}{8\pi}}\frac{x-iy}{r},\\
Y_{J=1,0}({\bm n})&=&\sqrt{\frac{3}{4\pi}}\frac{z}{r},
\end{eqnarray}
\end{subequations}
and the formula
\begin{align}\label{avang}
\int d\Omega\,n_in_j=\frac{4\pi}{3}\delta_{ij}
\end{align}
we can perform the integration over the angles and write the interaction Hamiltonian (\ref{pauli2}) in the form
\begin{align}\label{finham}
H_I=\sum_i{\bm\psi}^\dagger{\sigma}_i{\bm\psi}\!\int_0^\infty\!dk\,g(k)
\frac{c_i(k)+c_i^\dagger(k)}{\sqrt{2k}},
\end{align}
The annihilation and creation operators in the Cartesian basis are built from the operators (\ref{c}) as follows
\begin{eqnarray}\label{cart}
c_x(k)&=&\frac{c_-(k)-c_+(k)}{\sqrt{2}},\;\;
c_x^\dagger(k)=\frac{c_-^\dagger(k)-c_+^\dagger(k)}{\sqrt{2}},\nonumber\\
c_y(k)&=&\frac{c_-(k)+c_+(k)}{i\sqrt{2}},\;\;
c_y^\dagger(k)=i\frac{c_-^\dagger(k)+c_+^\dagger(k)}{\sqrt{2}},\nonumber\\
c_z(k)&=&c_0(k),\;\;c_z^\dagger(k)=c_0^\dagger(k).
\end{eqnarray}
They obey the standard commutation relations
\begin{eqnarray}\label{cra}
\left[c_i(k), c^{\dagger}_j(k')\right]=\delta_{ij}\delta(k-k').
\end{eqnarray}
The formfactor $g(k)$ is defined as
\begin{eqnarray}\label{formf0}
g(k)=-\frac{\mu k}{\pi\sqrt{3}}\int\!d^3r\,\rho'(r)j_1(kr).
\end{eqnarray}
The integral in this formula is proportional to the Fourier transform of $\rho(r)$
\begin{align}\label{formf1}
&-\int\!d^3r\,\rho'(r)j_1(kr)=-4\pi\int_0^\infty\!\!\!dr\rho'(r)r^2j_1(kr)\nonumber\\
&=4\pi\int_0^\infty\!\!\!dr \rho(r)\frac{d}{dr}
\left(\frac{\sin(kr)}{k^2}-r\frac{\cos(kr)}{k}\right)\nonumber\\
&=4\pi\int_0^\infty\!\!\!dr r\rho(r)\sin(kr)\nonumber\\
&=k\int\!d^3r\,e^{-i{\bm k}\cdot{\bm r}}\rho(r)=k{\tilde\rho}(k).
\end{align}

\subsection{Electric dipole coupling}

The calculation of the electric dipole Hamiltonian starts from the following formulas that are the counterparts of Eqs.~(\ref{hammag}):
\begin{subequations}\label{hamel}
\begin{align}
H_0&=\int\!d^3r\,{\bm\psi}^\dagger({\bm r})H_0^a{\bm\psi}({\bm r})\nonumber\\
&+\frac{1}{2}\int\!d^3r:\!\left({\bm E}^2({\bm r})+{\bm B}^2({\bm r})\right)\!:\,,\label{hamel0}\\
H_I&=-e\int\!d^3r\,{\bm\psi}^\dagger({\bm r}){\bm r}{\bm\psi}({\bm r})\!\cdot\!{\bm E}({\bm r}),\label{hamel1}
\end{align}
\end{subequations}
where $H_0^a$ is the atomic Hamiltonian. This time we will have four terms in the expansion of the electron field operators
\begin{align}\label{exped}
{\bm\psi}({\bm r})=\sum_i\frac{x^i}{r}\varphi_e(r){\bm\psi}_e^i+\varphi_g(r){\bm\psi}_g,
\end{align}
where we used the Cartesian basis for the three wave functions of the degenerate upper energy level. The wave functions of the excited states and the ground state appearing in this decomposition belong to the eigenvalues $m_e$ and $m_g$ of the atomic Hamiltonian. We continue to denote the energies by the letter $m$ to stress the analogy with the relativistic QED. The value of the dipole moment of the atomic transition between the ground state and an excited state, say the state described by $x\varphi_e(r)/r$, is
\begin{align}\label{d}
d=e\int\!d^3r\,\frac{x}{r}\varphi_e(r)x\varphi_g(r)=\frac{e}{3}\int\!d^3r\,\varphi_e(r)r\varphi_g(r).
\end{align}

Upon substituting the expansion (\ref{exped}) into Eqs.~(\ref{hamel}), we obtain
\begin{subequations}\label{hamelx}
\begin{align}
H_0&={\bm\psi}^\dagger {\breve m}{\bm\psi}+\frac{1}{2}\int\!d^3r:\!\left({\bm E}^2({\bm r})+{\bm B}^2({\bm r})\right)\!:\,,\label{hamelx0}\\
H_I&=-d{\bm\psi}^\dagger{\bm\tau}{\bm\psi}\!\cdot\!\int\!d^3r\,\kappa(r){\bm n}\left({\bm n}\!\cdot\!{\bm E}({\bm r})\right),\label{hamelx1}
\end{align}
\end{subequations}
where $\kappa(r)=e\varphi_e(r)r\varphi_g(r)/d$. Note, that according to Eq.~(\ref{d}), $\kappa(r)/3$ may be viewed as the normalized radial distribution function of the dipole moment. It plays the same role as $\rho(r)$ played in the description of the spin system. The diagonal matrix ${\breve m}$ describes the energy levels and the matrices ${\bm\tau}$ describe the transitions between the ground state and the excited states
\begin{align}\label{mat}
{\breve m}&=\left(\begin{array}{cccc}
m_e&0&0&0\\0&m_e&0&0\\0&0&m_e&0\\0&0&0&m_g\end{array}\right),\;\;
\tau_x=\left(\begin{array}{cccc}
0&0&0&1\\0&0&0&0\\0&0&0&0\\1&0&0&0\end{array}\right),\nonumber\\
\tau_y&=\left(\begin{array}{cccc}
0&0&0&0\\0&0&0&1\\0&0&0&0\\0&1&0&0\end{array}\right),\;\;
\tau_z=\left(\begin{array}{cccc}
0&0&0&0\\0&0&0&0\\0&0&0&1\\0&0&1&0\end{array}\right).
\end{align}
\begin{subequations}\label{ed}
Substituting the expansion into multipoles (\ref{dece}) of the electric field operator, we obtain two sets of integrals (and their complex conjugate counterparts)
\begin{align}
&\int\!d^3r\,\kappa(r){\bm n}\left({\bm n}\!\cdot\!{\bm E}_{JMk}^{(e)}({\bm r})\right)\nonumber\\ &=\int\!d^3r\,\kappa(r){\bm n}\left(i{\bm n}\!\cdot\!{\bm\nabla}{\bm\times}{\bm L}\,T_{JMk}({\bm r})\right),\label{eel}\\
&\int\!d^3r\,\kappa(r){\bm n}\left({\bm n}\!\cdot\!{\bm E}_{JMk}^{(m)}({\bm r})\right)\nonumber\\
&=\int\!d^3r\,\kappa(r){\bm n}\left({\bm n}\!\cdot\!{\bm L}\,kT_{JMk}({\bm r})\right).\label{emag}
\end{align}
\end{subequations}
All integrals in the second set vanish because ${\bm n}\!\cdot\!{\bm L}=0$. The integrals in the first set can be simplified with the use of the relations $i{\bm n}\!\cdot\!{\bm\nabla}{\bm\times}{\bm L}T_{JMk}({\bm r})=-{\bm L}^2T_{JMk}({\bm r})=-J(J+1)T_{JMk}({\bm r})$. We can again argue, as in our discussion of Eq.~(\ref{avb}), that the only nonvanishing contribution in the formula (\ref{eel}) comes from the dipole component, when $J=1$. Thus, the interaction Hamiltonian becomes
\begin{align}\label{ave1}
H_I&=d{\bm\psi}^\dagger{\bm\tau}{\bm\psi}\!\cdot\!\int\!d^3r\,\sum_M\int_0^{\infty}
dk\,\sqrt{\frac{2k}{\pi}}j_1(kr)\kappa(r)\nonumber\\
&\times {\bm n}\left[Y_{1M}({\bm n})d_M(k)+Y_{1M}^*({\bm n})d_M^\dagger(k)\right],
\end{align}
where we have introduced again a simplified notation for the annihilation and creation operators
\begin{align}\label{ce}
d_+(k)=c_{1,1}^{(e)}(k),\;\;\;d_0(k)=c_{1,0}^{(e)}(k),\;\;\;d_-(k)=c_{1,-1}^{(e)}(k).
\end{align}
Note, that the electric dipole Hamiltonian (\ref{ave1}) has the same general form as the magnetic dipole Hamiltonian (\ref{avb1}). Therefore, we may use the same methods to transform (\ref{ave1}) to the form
\begin{align}\label{finhame}
H_I={\bm\psi}^\dagger{\tau}_i{\bm\psi}\!\int_0^\infty\!dk\,{\breve g}(k)\frac{d_i(k)+d_i^\dagger(k)}{\sqrt{2k}},
\end{align}
where
\begin{align}\label{ge}
{\breve g}(k)=\frac{d k}{\pi\sqrt{3}}\int\!d^3r\,\kappa(r)j_1(kr).
\end{align}
\widetext

\section{Radiative corrections to the photon propagator}\label{a2}

\subsection{Spin system}

The contributions to the photon self-energy in the fourth-order of perturbation theory corresponding to the diagrams (b), (c), (d), and (e) in Fig.~\ref{Fig6} lead to the following integrals:
\begin{align}
{\cal P}_{ab}^{(4b)}(k_0) &= \int_{-\infty}^{\infty}\!\frac{d p_0}{2\pi}\int_{-\infty}^{\infty}\!\frac{d l_0}{2\pi}\mathrm{Tr}\left\{\sigma_aS_F(p_0+k_0)\sigma_m S_F(p_0+k_0+l_0)\sigma_b S_F(p_0+l_0)\sigma_n S_F(p_0)\right\}D_{Fmn}(l_0),\\
{\cal P}_{ab}^{(4c)}(k_0) &= \int_{-\infty}^{\infty}\!\frac{d p_0}{2\pi}\int_{-\infty}^{\infty}\!\frac{d l_0}{2\pi}\mathrm{Tr}\left\{\sigma_aS_F(p_0+k_0)\sigma_b S_F(p_0)\sigma_m S_F(p_0+l_0)\sigma_n S_F(p_0)\right\}D_{Fnm}(l_0),\\
{\cal P}_{ab}^{(4d)}(k_0) &= i\,\delta m_s\!\int_{-\infty}^{\infty}\!\frac{d p_0}{2\pi} \mathrm{Tr}\left\{\sigma_aS_F(p_0+k_0)\sigma_b S_F(p_0)\sigma_z S_F(p_0)\right\}\\
{\cal P}_{ab}^{(4e)}(k_0) &= -i\,m_t\!\int_{-\infty}^{\infty}\!\frac{d p_0}{2\pi} \mathrm{Tr}\left\{\sigma_aS_F(p_0+k_0)\sigma_b S_F(p_0)\sigma_z S_F(p_0)\right\}.
\end{align}
The contributions corresponding to the diagrams (f), (g), and (h) are the same as those corresponding to the diagrams (c), (d), and (e). Therefore, we shall calculate the contributions only from the diagrams (c), (d), and (e) and multiply them by 2. Again, as in the lowest order, the matrices ${\cal P}_{ab}$ are diagonal in the angular momentum basis. All Feynman integrals can be evaluated by the method of residues. We shall present the detailed calculation of the first integral. The three remaining contributions are even simpler to calculate and we shall give only the final results. The nonvanishing components of ${\cal P}_{ab}^{(4b)}(k_0)$ in the angular momentum basis are ${\cal P}_{\pm}^{(4b)}(k_0)$ and ${\cal P}_{0}^{(4b)}(k_0)$
\begin{subequations}
\begin{align}
&{\cal P}_{+}^{(4b)}(k_0) = \int_0^\infty\!dk\,g^2(k)\int_{-\infty}^{\infty}\!\frac{d p_0}{2\pi}\int_{-\infty}^{\infty}\!\frac{d l_0}{2\pi}\frac{1}{l_0^2-k^2+i\epsilon}\nonumber\\
&\times\mathrm{Tr}\left\{\sigma_-\frac{1}{p_0+k_0-(m-i\epsilon)\sigma_z}\sigma_z
\frac{1}{p_0+k_0+l_0-(m-i\epsilon)\sigma_z}\sigma_+ \frac{1}{p_0+l_0-(m-i\epsilon)\sigma_z}\sigma_z\frac{1}{p_0-(m-i\epsilon)\sigma_z}\right\}\\
&= -2\int_0^\infty\!dk\,g^2(k)\int_{-\infty}^{\infty}\!\frac{d p_0}{2\pi}\int_{-\infty}^{\infty}\!\frac{d l_0}{2\pi}\frac{1}{l_0^2-k^2+i\epsilon}
\frac{1}{p_0+k_0-m+i\epsilon}\frac{1}{p_0+k_0+l_0-m+i\epsilon} \frac{1}{p_0+l_0+m-i\epsilon}\frac{1}{p_0+m-i\epsilon},\nonumber\\
&{\cal P}_{-}^{(4b)}(k_0) = \int_0^\infty\!dk\,g^2(k)\int_{-\infty}^{\infty}\!\frac{d p_0}{2\pi}\int_{-\infty}^{\infty}\!\frac{d l_0}{2\pi}\frac{1}{l_0^2-k^2+i\epsilon}\nonumber\\
&\times\mathrm{Tr}\left\{\sigma_+\frac{1}{p_0+k_0-(m-i\epsilon)\sigma_z}\sigma_z
\frac{1}{p_0+k_0+l_0-(m-i\epsilon)\sigma_z}\sigma_- \frac{1}{p_0+l_0-(m-i\epsilon)\sigma_z}\sigma_z\frac{1}{p_0-(m-i\epsilon)\sigma_z}\right\}\\
&= -2\int_0^\infty\!dk\,g^2(k)\int_{-\infty}^{\infty}\!\frac{d p_0}{2\pi}\int_{-\infty}^{\infty}\!\frac{d l_0}{2\pi}\frac{1}{l_0^2-k^2+i\epsilon}
\frac{1}{p_0+k_0+m-i\epsilon}\frac{1}{p_0+k_0+l_0+m-i\epsilon} \frac{1}{p_0+l_0-m+i\epsilon}\frac{1}{p_0-m+i\epsilon},\nonumber\\
&{\cal P}_{0}^{(4b)}(k_0) = \int_0^\infty\!dk\,g^2(k)\int_{-\infty}^{\infty}\!\frac{d p_0}{2\pi}\int_{-\infty}^{\infty}\!\frac{d l_0}{2\pi}\frac{1}{l_0^2-k^2+i\epsilon}\nonumber\\
&\times\left[\mathrm{Tr}\left\{\sigma_z\frac{1}{p_0+k_0-(m-i\epsilon)\sigma_z}\sigma_+
\frac{1}{p_0+k_0+l_0-(m-i\epsilon)\sigma_z}\sigma_z \frac{1}{p_0+l_0-(m-i\epsilon)\sigma_z}\sigma_-\frac{1}{p_0-(m-i\epsilon)\sigma_z}\right\}\right.\nonumber\\
&+\left.\mathrm{Tr}\left\{\sigma_z\frac{1}{p_0+k_0-(m-i\epsilon)\sigma_z}\sigma_-
\frac{1}{p_0+k_0+l_0-(m-i\epsilon)\sigma_z}\sigma_z \frac{1}{p_0+l_0-(m-i\epsilon)\sigma_z}\sigma_+\frac{1}{p_0-(m-i\epsilon)\sigma_z}\right\}\right]\\
&= -2\int_0^\infty\!dk\,g^2(k)\int_{-\infty}^{\infty}\!\frac{d p_0}{2\pi}\int_{-\infty}^{\infty}\!\frac{d l_0}{2\pi}\frac{1}{l_0^2-k^2+i\epsilon}\frac{1}{p_0+k_0-m+i\epsilon}\frac{1}{p_0+k_0+l_0+m-i\epsilon} \frac{1}{p_0+l_0+m-i\epsilon}\frac{1}{p_0-m+i\epsilon}\nonumber\\
&-2\int_0^\infty\!dk\,g^2(k)\int_{-\infty}^{\infty}\!\frac{d p_0}{2\pi}\int_{-\infty}^{\infty}\!\frac{d l_0}{2\pi}\frac{1}{l_0^2-k^2+i\epsilon}\frac{1}{p_0+k_0+m-i\epsilon}\frac{1}{p_0+k_0+l_0-m+i\epsilon} \frac{1}{p_0+l_0-m+i\epsilon}\frac{1}{p_0+m-i\epsilon}\nonumber.
\end{align}
\end{subequations}
Note the advantage of using the angular momentum bases --- in each case we are left with only one or two integrals. We dropped several terms that did not contribute because they contained the squares of the $\sigma_\pm$ matrices. Finally, we replaced all $\sigma_z$ matrices by $\pm 1$ due to the presence of the projection operators ${\mathbb P}_e$ and ${\mathbb P}_g$. The results of the integrations are
\begin{subequations}
\begin{align}
{\cal P}_{+}^{(4b)}(k_0) &= \frac{2}{(2m-k_0)^2}\int_0^\infty\!\frac{dk}{k}\frac{g^2(k)}{k+2m-k_0-i\epsilon},\\
{\cal P}_{-}^{(4b)}(k_0) &= {\cal P}_{+}^{(4b)}(-k_0),\\
{\cal P}_{0}^{(4b)}(k_0) &= -4\int_0^\infty\!\frac{dk}{k}\frac{g^2(k)}{k+2m}\,\frac{1}{(k+2m)^2-k_0^2-i\epsilon}.
\end{align}
\end{subequations}
The contributions corresponding to the diagrams (c), (d), and (e) are
\begin{subequations}
\begin{align}
{\cal P}_{+}^{(4c)}(k_0) &= \frac{2}{2m-k_0}\int_0^\infty\!\frac{dk}{k}\frac{g^2(k)}{(k+2m)^2}
-\frac{1}{(2m-k_0)^2}\int_0^\infty\!\frac{dk}{k}g^2(k)\left(\frac{1}{k+2m-k_0-i\epsilon}
-\frac{2}{k+2m}\right),\\
{\cal P}_{-}^{(4c)}(k_0) &= {\cal P}_{+}^{(c)}(-k_0),\;\;\;
{\cal P}_{0}^{(4c)}(k_0)
= -2\int_0^\infty\!\frac{dk}{k}\frac{g^2(k)}{k+2m}\frac{1}{(k+2m)^2-k_0^2-i\epsilon},\\
{\cal P}_{+}^{(4d)}(k_0) &= -\frac{2\,\delta m_s}{(2m-k_0)^2},\;\;\;
{\cal P}_{-}^{(4d)}(k_0) = {\cal P}_{+}^{(4d)}(-k_0),\;\;\;
{\cal P}_{0}^{(4d)}(k_0) = 0\\
{\cal P}_{+}^{(4e)}(k_0) &= \frac{2 m_t}{(2m-k_0)^2},\;\;\;
{\cal P}_{-}^{(4e)}(k_0) = {\cal P}_{+}^{(4e)}(-k_0),\;\;\;
{\cal P}_{0}^{(4e)}(k_0) = 0.
\end{align}
\end{subequations}
Collecting all contributions in the second and fourth order, we obtain the following formulas for the three angular momentum components of the photon self-energy part:
\begin{subequations}\label{all}
\begin{align}
{\cal P}_{\pm}^{(2+4)}(k_0)&=-2(1-b)\left[\frac{1}{2m\mp k_0}
+\frac{\delta}{(2m\mp k_0)^2}\right],\label{all1}\\
{\cal P}_{0}^{(2+4)}(k_0)
&=-4\int_0^\infty\!\frac{dk}{k}\frac{g^2(k)}{k+2m}\,\frac{1}{(k+2m)^2-k_0^2-i\epsilon},\label{all3}
\end{align}
\end{subequations}
where
\begin{align}
b=2\int_0^\infty\!\frac{dk}{k}\frac{g^2(k)}{(k+2m)^2},\quad
\delta=\frac{1}{2(1-b)}\int_0^\infty\!\frac{dk\,g^2(k)}{k^2}.
\end{align}

\subsection{Two-level atom}

The calculations of the fourth-order corrections to the photon propagator for the two-level atom are much simpler than those for the spin system. First, the propagator has only one component. Second, the contributions corresponding to the tadpole diagrams vanish. Third, there is no summation over three different states in the internal photon lines. The contribution corresponding to the diagram (b) in  Fig.~\ref{Fig6} is
\begin{align}
&{\cal\hat P}^{(4b)}(k_0) = \int_0^\infty\!dk\,g^2(k)\int_{-\infty}^{\infty}\!\frac{d p_0}{2\pi}\int_{-\infty}^{\infty}\!\frac{d l_0}{2\pi}\frac{1}{l_0^2-k^2+i\epsilon}\nonumber\\
&\times\mathrm{Tr}\left\{\sigma_x\frac{1}{p_0+k_0-(m-i\epsilon)\sigma_z}\sigma_x
\frac{1}{p_0+k_0+l_0-(m-i\epsilon)\sigma_z}\sigma_x \frac{1}{p_0+l_0-(m-i\epsilon)\sigma_z}\sigma_x\frac{1}{p_0-(m-i\epsilon)\sigma_z}\right\}\\
&= -\int_0^\infty\!dk\,g^2(k)\int_{-\infty}^{\infty}\!\frac{d p_0}{2\pi}\int_{-\infty}^{\infty}\!\frac{d l_0}{2\pi}\frac{1}{l_0^2-k^2+i\epsilon}
\frac{1}{p_0+k_0+m-i\epsilon}\frac{1}{p_0+k_0+l_0-m+i\epsilon} \frac{1}{p_0+l_0+m-i\epsilon}\frac{1}{p_0-m+i\epsilon}\nonumber\\
&+\int_0^\infty\!dk\,g^2(k)\int_{-\infty}^{\infty}\!\frac{d p_0}{2\pi}\int_{-\infty}^{\infty}\!\frac{d l_0}{2\pi}\frac{1}{l_0^2-k^2+i\epsilon}
\frac{1}{p_0+k_0-m+i\epsilon}\frac{1}{p_0+k_0+l_0+m-i\epsilon} \frac{1}{p_0+l_0-m+i\epsilon}\frac{1}{p_0+m-i\epsilon}.
\end{align}
Similar integrals are obtained for the diagrams (c) and (e) in Fig.~\ref{Fig6}. The results of the integrations are
\begin{subequations}
\begin{align}
{\cal\hat P}^{(4b)}(k_0) &= \frac{2}{4m^2-k_0^2}\int_0^\infty\!\frac{dk}{k}\frac{\hat{g}^2(k)}{k+2m},\\
{\cal\hat P}^{(4c)}(k_0) &= -\frac{1}{4m^2-k_0^2}\int_0^\infty\!dk\frac{\hat{g}^2(k)}{(k+2m)^2}
-\frac{8m^2}{(4m^2-k_0^2)^2}\int_0^\infty\!\frac{dk}{k}\frac{\hat{g}^2(k)}{k+2m},\\
{\cal\hat P}^{(4d)}(k_0) &= \delta m_a\left(\frac{2}{4m^2-k_0^2}+\frac{16m^2}{(4m^2-k_0^2)^2}\right).
\end{align}
\end{subequations}
The sum of the contributions from all diagrams in the second and fourth-order is
\begin{align}\label{pse2l}
{\cal\hat P}^{(2+4)}(k_0) = {\cal\hat P}^{(2)}(k_0)+{\cal\hat P}^{(4b)}(k_0)+2{\cal\hat P}^{(4c)}(k_0)
+2{\cal\hat P}^{(4d)}(k_0)= -\frac{4m(1-\hat{b})}{4m^2-k_0^2},
\end{align}
where
\begin{align}
{\hat b} = \frac{1}{2m}\int_0^\infty\!\frac{dk}{k}\frac{\hat{g}^2(k)(k+4m)}{(k+2m)^2}.
\end{align}

\subsection{Electric dipole atom}

For completeness, we also present the self-energy parts in the fourth order for the dipole atom. They are not much different from those for the two-level atom and we list here only the final results. There is one difference that is worth mentioning. Namely, the contributions corresponding to the diagrams (c) and (d) are now different from the contributions corresponding to the diagrams (f) and (g). This difference is due to the triple degeneracy of the excited energy level that breaks the symmetry between the $|e\rangle$ and $|g\rangle$ states that existed for the spin system and the two-level atom. Again, there are no contributions from tadpole diagrams. The self-energy parts in the fourth order are
\begin{subequations}
\begin{align}
{\cal\breve P}_{ij}^{(4b)}(k_0) &= \delta_{ij}\frac{2}{\Delta m^2-k_0^2}\int_0^\infty\!\frac{dk}{k}\frac{\breve{g}^2(k)}{k+\Delta m},\\
{\cal\breve P}_{ij}^{(4c)}(k_0)+{\cal\breve P}_{ij}^{(4d)}(k_0) &= \delta_{ij}\frac{2\Delta m+k_0}{\Delta m^2-k_0^2}\int_0^\infty\!dk\frac{\breve{g}^2(k)}{k(k+\Delta m)^2},\\
{\cal\breve P}_{ij}^{(4f)}(k_0)+{\cal\breve P}_{ij}^{(4g)}(k_0) &= \delta_{ij}\frac{2\Delta m-k_0}{\Delta m^2-k_0^2}\int_0^\infty\!dk\frac{\breve{g}^2(k)}{k(k+\Delta m)^2},
\end{align}
\end{subequations}
where we used the values (\ref{mcorr}) for the mass corrections. The total self-energy part in the second and fourth order is (disregarding the Kronecker symbols)
\begin{align}\label{pseda}
{\cal\breve P}^{(2+4)}(k_0) = {\cal\breve P}^{(2)}(k_0)+{\cal\breve P}^{(4b)}(k_0)+{\cal\breve P}^{(4c)}(k_0)+{\cal\breve P}^{(4d)}(k_0)
+{\cal\breve P}^{(4f)}(k_0)+{\cal\breve P}^{(4g)}(k_0)= -\frac{2\Delta m(1-\breve{b})}{\Delta m^2-k_0^2},
\end{align}
where
\begin{align}
{\breve b} = \frac{1}{\Delta m}\int_0^\infty\!\frac{dk}{k}\frac{\breve{g}^2(k)(k+3\Delta m)}{(k+\Delta m)^2}.
\end{align}

\end{document}